\documentclass[12pt,pra,aps,amssymb,amsfonts,amsmath,tightenlines]{revtex4}%aps instead of rmp
\usepackage{dsfont}
\usepackage{graphicx}
\usepackage{amssymb,amsfonts,amsthm}
\usepackage{color}
\usepackage{verbatim}
\usepackage[normalem]{ulem}
\usepackage{enumerate}
\usepackage{mathrsfs}
\usepackage{bm,float}

\usepackage{textgreek}
\usepackage{lgreek}

\newtheorem{Proposition}{Proposition}
\usepackage{nicefrac}
\newcommand{\Var}{\text{Var}}

\newcommand{\half}{\mbox{$\textstyle \frac{1}{2}$}}
\newcommand{\quat}{\mbox{$\textstyle \frac{1}{4}$}}
\newcommand{\re}{\mbox{$\rm e$}}
\newcommand{\ri}{\mbox{$\rm i$}}
\newcommand{\rd}{\mbox{$\rm d$}}
\newcommand{\octa}{\mbox{$\textstyle \frac{1}{8}$}}

\begin{document}

\title{L\'evy Models for Collapse of the Wave Function}
\author{Dorje C.~Brody$^{1,2}$ and Lane~P.~Hughston$^3$}

\affiliation{
$^1$Department of Mathematics, University of Surrey\\
Guildford GU2 7XH, United Kingdom\\
$^2$St Petersburg National Research University of Information Technologies, Mechanics and Optics, St Petersburg 197101, Russia\\
$^3$Department of Computing, Goldsmiths University of London\\ New Cross, London SE14\,6NW, United Kingdom
}

\date{\today}
%ABSTRACT HERE
\begin{abstract}
\noindent 
Recently there has been much progress in the development of stochastic models
for state reduction in quantum mechanics.  In such models, the collapse of the wave function is 
a physical process, governed by a nonlinear stochastic differential equation that generalizes
the Schr\"odinger equation. The present paper considers energy-based stochastic extensions of the Schr\"odinger equation. 
Most of the work carried out hitherto in this area has been concerned with models where the process driving the stochastic dynamics of the quantum state is  Brownian motion. Here, the Brownian framework is broadened to a wider class of models where the noise process is of the L\'evy type, admitting stationary and independent increments. The properties of such models are different from those of Brownian reduction models. In particular, for L\'evy models the decoherence rate depends on the overall scale of the energy. Thus, in L\'evy  reduction models, a macroscopic quantum system will spontaneously collapse to an eigenstate even if the energy level gaps are small.
\vspace{-0.2cm}
\\
%KEY WORDS HERE
\begin{center}
\end{center}
%\keywords{Key Words}
%\subclass{MSC code\and JEL classification code}
\end{abstract}

\maketitle
%SECTION
%%%%%%%%%%%%%%%
\section{The stochastic Schr\"odinger equation}
\label{sec:1} 

A number of authors have worked on the development of 
dynamical models for the collapse of the wave function  \cite{ah2000, ab2001, abbh2001, adler2002, adler2003, bh2002-2, bh2005, bh2018, diosi1988a, diosi1988b, diosi1989, grw1986,  gpr1990, gisin1989, gisinpercival1993, hughston1996, pearle1989, percival1994, pearle1, weinberg2012}. 
For overviews see \cite{bg2003, bh2006, blssu2013, guo}. Such models have a highly nontrivial relationship with the probabilistic hypotheses
of standard quantum mechanics. Progress in this area can be
classified into work on (a) spontaneous localization of the state and (b) collapse of the state vector to an energy
eigenstate. We are concerned with the latter here. Our goal is to show how the well-established framework for stochastic state reduction based on Brownian noise can be extended to a wider class of models based on noise processes with stationary and independent increments, so-called L\'evy processes. In general, such processes have jumps. A  L\'evy process is continuous if and only if it is a Brownian motion. A pure jump L\'evy process can be decomposed into the sum of a finite activity process and an infinite activity process. Processes of finite activity have the property that jumps occur at a finite rate. Processes of infinite activity jump infinitely often over any finite interval of time. 
We provide examples of state reduction models based on each of these  types of L\'evy processes. We argue that there is no reason {\em a priori} to prefer continuous processes over discontinuous processes in models for quantum state reduction. 
\newpage
For the dynamics of the state vector in the simplest energy-driven model with a Brownian driver, we have the following well-known stochastic differential equation of the Ito type defined on a finite dimensional Hilbert space:
\begin{eqnarray} \label{stochastic Schrodinger equation}
{\rm d}|\psi_t\rangle = -{\rm i}\hbar^{-1}{\hat H} |\psi_t\rangle {\rm d}t -
\octa \sigma^2({\hat H}-H_t)^2 |\psi_t\rangle {\rm d}t + \half
\sigma ({\hat H}-H_t)|\psi_t\rangle {\rm d}W_t .
\end{eqnarray}
Here $|\psi_t\rangle$ denotes the state at time $t$, with initial condition $|\psi_0\rangle$, ${\hat H}$ is
the Hamiltonian, $\{W_t\}_{t\geq0}$ is a standard one-dimensional
Brownian motion, and
\begin{eqnarray} \label{energy expectation}
H_t = \frac{\langle{\psi}_t|{\hat H}|\psi_t\rangle}
{\langle{\psi}_t|\psi_t\rangle}  
\end{eqnarray}
is the expectation of ${\hat H}$ in the state $|\psi_t\rangle$. 
The parameter $\sigma$, which has the units 
\begin{eqnarray} \label{units}
\sigma\sim[{\rm
energy}]^{-1}[{\rm time}]^{-1/2},
\end{eqnarray}
determines the characteristic
timescale $ \tau_R = 1/\sigma^2 V_0$ for the collapse of the wave function. Here $V_0$ denotes the initial value of the squared uncertainty of the energy. More generally, the conditional variance of the energy is defined for $t\geq 0$ by 
\begin{eqnarray} \label{energy variance}
V_t = \frac{\langle{\psi}_t|({\hat H}-H_t)^2|\psi_t\rangle}
{\langle{\psi}_t|\psi_t\rangle}.  \label{eq:3}
\end{eqnarray}
The energy-conserving stochastic  Schr\"odinger equation based on Brownian 
driver, originally introduced in  \cite{gisin1989}, is the simplest known rigorous model for state reduction in which the Born probability rules can be derived dynamically.

The dynamics of $|\psi_t\rangle$ set out in \eqref{stochastic Schrodinger equation} are defined on a probability space
$(\Omega,{\mathcal F},{\mathbb P})$ with filtration $\{{\mathcal
F}_t\}_{0\leq t<\infty}$. For convenience we recall some key definitions \cite{hughston1996, abbh2001, bh2002-2, bh2006}. The elements of $\Omega$ represent the possible outcomes of chance in the model under consideration.  The event space ${\mathcal F}$ is a $\sigma$-algebra of subsets of $\Omega$. The measure $\mathbb P$ assigns a probability $\mathbb P : A \in {\mathcal F} \mapsto \mathbb P(A) \in [0, 1]$ to each event $A$ in 
${\mathcal F}$, given by
\begin{eqnarray}
{\mathbb P} (A) = \int_{\Omega} \mathds1 _A(\omega)\, \mathbb P(\rd \omega),
\end{eqnarray}
where  $\mathds1_A$ denotes the indicator function of the set $A\subset \Omega$, taking the value one if $\omega\in A$, and zero otherwise. Here $\omega \in \Omega$ denotes a typical outcome of chance.
A function $X: \Omega \to \mathbb R$ is said to be a random variable on $(\Omega,{\mathcal F},{\mathbb P})$ if $X$ is ${\mathcal F}$-measurable, that is, if for any $x \in \mathbb R \cup\{\pm \infty\}$ it holds that the set $\{ \omega \in \Omega : X(\omega) \leq x \}$ is an element of ${\mathcal F}$. The distribution of $X$ is the function $F_X:\mathbb R \cup\{\pm \infty\} \to [0,1]$ defined by $F_X(x) = \mathbb P(X \leq x)$. The expectation of $X$, which takes values in the extended real line, is then defined by the Lebesgue integral 
\begin{eqnarray}
{\mathbb E} [X] = \int_{\Omega} X(\omega)\, \mathbb P(\rd \omega).
\end{eqnarray}
We note that if 
${\mathbb E} [ \max(X,0)] = \infty$ and ${\mathbb E} [ \min(X,0)] =- \infty$ then ${\mathbb E} [X]$ is not defined. We say that $X$ is integrable under $\mathbb P$ if ${\mathbb E} [\,|X|\,] < \infty$.
The subtle measure-theoretic definition of  conditional expectation due to Kolmogorov  \cite{kolmogorov1933} involves a construction  that generalizes the elementary notion of conditional probability and adds more precision to the idea. 
Specifically, if ${\mathcal E}$ is a sub-$\sigma$-algebra of the $\sigma$-algebra
${\mathcal F}$ on a probability space $(\Omega,{\mathcal F},{\mathbb P})$ and if $X$ is an integrable random variable, then the  conditional
expectation ${\mathbb E}[X|{\mathcal E}]$ of $X$ with respect to ${\mathcal
E}$ is defined as follows. We write 
$Y =  {\mathbb E}[X|{\mathcal E}]$ for any ${\mathcal E}$-measurable random variable $Y$ with the property that for any $A \in {\mathcal E}$ it holds that
\begin{eqnarray}
\int_{\Omega} \mathds1_A(\omega) X(\omega)\, \mathbb P(\rd \omega) =
\int_{\Omega}  \mathds1_A(\omega) Y(\omega)\, \mathbb P(\rd \omega).
\end{eqnarray}
The conditional expectation is unique modulo differences on sets of $\mathbb P$-measure zero. Any particular choice of 
$Y$ from such an equivalence class is called a version of ${\mathbb E}[X|{\mathcal E}]$.  
Then if ${\mathcal D}$ is a sub-$\sigma$-algebra of ${\mathcal E}$, and  ${\mathcal E}$ is a sub-$\sigma$-algebra of ${\mathcal F}$, we have the tower property of conditional expectation: 
$ {\mathbb E} [ {\mathbb E}[X|{\mathcal E}] | {\mathcal D}] = {\mathbb E}[X|{\mathcal D}]$. 
In particular, because the trivial $\sigma$-algebra 
${\mathcal Z} = \{\Omega, \emptyset\}$ satisfies ${\mathbb E}[X|{\mathcal Z}] = {\mathbb E}[X]$ for any integrable random variable $X$, and ${\mathcal Z}$ is a sub-$\sigma$-algebra of ${\mathcal E}$, we have 
$ {\mathbb E} [ {\mathbb E}[X|{\mathcal E}]] = {\mathbb E}[X]$ for any sub-$\sigma$-algebra ${\mathcal E}$.

The filtration $\{{\mathcal F}_t\}_{0\leq t<\infty}$ consists of a nested family of 
sub-$\sigma$-algebras of ${\mathcal F}$ such that $s\leq t$ implies that 
${\mathcal F}_s$ is a sub-$\sigma$-algebra of ${\mathcal F}_t$. In the case of 
a filtration we often use the simplifying notation ${\mathbb E}_t[X]=
{\mathbb E}[X|{\mathcal F}_t]$ for the conditional expectation. Then by the 
tower property we have ${\mathbb E}_s[{\mathbb E}_t[X]]={\mathbb E}_s[X]$ 
for $s\leq t$. 

By a random process (or stochastic process) we mean a collection of random variables 
$\{X_t\}_{t\geq 0}$. A random process is said  to be adapted to 
$\{{\mathcal F}_t\}$ if for each $t \geq 0$ it holds that $X_t$ is ${\mathcal
F}_t$-measurable. An adapted process $\{X_t\}_{t\geq 0}$ is said to be a
martingale if ${\mathbb E}[|X_t|]<\infty$ for $t \geq 0$ and ${\mathbb
E}_s[X_t]=X_s$ for $0\leq s\leq t<\infty$.  Thus the martingale property characterizes the dynamics of a quantity that at each step is only on 
average conserved. A process $\{X_t\}_{t\geq 0}$ is a supermartingale
if ${\mathbb E}[|X_t|]<\infty$ for $t \geq 0$ and ${\mathbb E}_s[X_t]\leq X_s$ 
for all $0\leq s\leq t<\infty$. 

With these definitions at hand one finds as a consequence of \eqref{stochastic Schrodinger equation} that the expectation of the energy is a martingale and that the
variance of the energy is a supermartingale. That is to say, we have
\begin{eqnarray}
{\mathbb E}_s[H_t]=H_s \, , \qquad {\mathbb E}_s[V_t] \leq V_s .
\end{eqnarray}
These relations can be worked out by an application of Ito's lemma to
(\ref{energy expectation}) and (\ref{energy variance}), from which one infers
\begin{eqnarray}
{\rm d}H_t = \sigma V_t \,{\rm d}W_t , \qquad
{\rm d}V_t=-\sigma^2 V_t^2\,{\rm d}t+\sigma \kappa_t \,{\rm d}W_t ,
\end{eqnarray}
where $\kappa_t =  \langle{\psi}_t|({\hat H}-H_t)^3
|\psi_t\rangle / \langle{\psi}_t|\psi_t\rangle. $
In particular, since the energy is bounded, the fact that $\{H_t\}$ has no drift implies that it is a
martingale. Then the fact that the drift of $\{V_t\}$ is negative
shows that $\{V_t\}$ is a supermartingale.

The martingale condition can thus be interpreted as the form of the energy conservation law 
that applies even when a system is not in a definite state of energy. 
The supermartingale property satisfied by $\{V_t\}$ captures the essence of
what is meant by a  reduction process in quantum mechanics. In particular, one can show by use of \eqref{stochastic Schrodinger equation} that 
\begin{eqnarray}
\lim_{t\to\infty}{\mathbb E}\left[V_t\right] = 0,
\end{eqnarray}
which implies that reduction proceeds to an energy eigenstate, because 
only at an energy eigenstate do we have $V_t=0$. Writing $H_\infty$ for the 
random terminal value of the energy, one can prove that
$ H_t = {\mathbb E}_t[H_\infty]$ and
$V_t = {\mathbb E}_t[(H_\infty-H_t)^2]$.
That is to say, $\{H_t\}$ and $\{V_t\}$ are
respectively the ${\mathcal F}_t$-conditional mean and variance of
the terminal value of the energy after reduction. In particular, we have the relations
\begin{eqnarray}
H_0 = {\mathbb E}[H_\infty], \qquad
V_0 = {\mathbb E}[(H_\infty-H_0)^2],
\end{eqnarray}
which form the basis of the statistical interpretation of quantum mechanics \cite{hughston1996, ah2000, abbh2001, bh2002-2}. The first of these  shows that the so-called expectation value of the observable $\hat H$ in the state $|\psi_0\rangle$ is equal to the expectation (in the probabilistic sense) of the random variable corresponding to the outcome of the measurement of the energy. 
Similarly, the squared uncertainty of $\hat H$ in the state $|\psi_0\rangle$ is equal to the variance of the outcome of the measurement. These results all carry through to the L\'evy-based models that we consider shortly.

%%%%%%%%%%%%%%%%%%%%%
\section{Generalization to Mixed States}
\label{sec:2}
%%%%%%%%%%%%%%%%%%%%%

If we take the view that the general state of a quantum system is described by a density matrix, and that state reduction prevails at the 
level of density matrices, then the dynamical equation for the reduction of the 
density matrix $\{{\hat\rho}_t\}_{t \geq0}$ takes the form of the following stochastic master equation \cite{bh2018}:
\begin{eqnarray}\label{density matrix dynamics}
\rd{\hat\rho}_t = -{\rm i} \hbar^{-1} [{\hat H},{\hat\rho}_t]\, \rd t +\tfrac{1}{4} \sigma^2
{\mathcal L}_{\hat H}\,{\hat\rho}_t \,\rd t  + \half\sigma \big( ({\hat
H}-H_t){\hat\rho}_t + {\hat\rho}_t({\hat H}-H_t) \big) \rd W_t.
\end{eqnarray}
Here we write
$
H_t={\rm tr} ( {\hat H}{\hat\rho}_t )
$
for the expectation of the Hamiltonian 
and ${\mathcal L}_{\hat H}$ denotes the Lindblad-GKS super-operator \cite{lindblad, GKS}, given in the present context by
\begin{eqnarray}
{\mathcal L}_{\hat H}\,{\hat\rho}_t = {\hat H} {\hat\rho}_t {\hat
H} -\half {\hat\rho}_t {\hat H}^2 - \half {\hat H}^2 {\hat\rho}_t .
\end{eqnarray}

One can show that the dynamical equation for ${\hat\rho}_t$ has the following properties: (a) the trace of ${\hat\rho}_t$ is preserved, 
(b) the positivity of ${\hat\rho}_t$ is preserved, 
(c) the conditional expectation of the energy has the martingale property 
(energy is conserved on average), and (d) the variance
of the energy is a supermartingale (state reduction occurs). 
Additionally,   we find that if $\hat \Pi_j$ denotes the projection operator onto the Hilbert subspace associated to
the energy eigenvalue $E_j$, then (e) the process $\{\Pi_{jt} \}_{t \geq 0}$ defined by
\begin{eqnarray}
\Pi_{j t} = {\rm tr} \, (  \hat \rho_t \,\hat \Pi_{j} )
\end{eqnarray}
is a martingale. This generalizes a result of Adler \& Horwitz \cite{ah2000} and shows the Born rule can be deduced under the dynamical equation \eqref{density matrix dynamics} in the situation where the system is in a 
random mixed state. One can see at a glance that the mean density matrix will satisfy the Lindblad equation, a property that is not so obvious from the dynamics of the state vector. In fact, in our dynamical model for the density matrix we find that state reduction proceeds in accordance with the L\"uders projection postulate \cite{luders1951} to the L\"uders eigenstate associated with the eigenvalue $E_j$ obtained as the result of the measurement: 
\begin{eqnarray}
\hat \rho_0 \to \frac {\hat \Pi_j \,  \hat \rho_0 \, \hat \Pi_j} { {\rm tr} \, ( \hat \rho_0 \, \hat \Pi_{j}  ) }.
\end{eqnarray}

It is worth emphasizing, in this connection,  that the Born rule is an {\it assumption} in standard quantum mechanics, part of the statistical interpretation of the theory. The same holds for the L\"uders projection postulate. 
But the Born rule and the L\"uders projection postulate both arise as {\it theorems} of the energy-driven reduction model, even for initially mixed states. 
In this respect, the energy-driven model can be regarded as superior to (a) the GRW model \cite{grw1986} and (b) the CSL (continuous spontaneous localization) model \cite{pearle1989, gpr1990}, neither of which exhibit this property on an exact basis. 

Now, one might ask whether it even makes sense to speak of the reduction of a mixed state. After all, a mixed state is usually understood to represent an ensemble, rather than a single particle. Indeed, most of the literature of dynamic state reduction models adheres to such a view. It seems that early on, since the work of \cite{grw1986, pearle1989, gisin1990, gpr1990}, the idea that a single quantum system in isolation is necessarily represented by a pure state was somehow embedded in the thinking of most of those who were working on reduction models. It may be that this view simply echoed the thinking of the majority of physicists at the time, and it is not surprising that such a view continues to be held by many to this day. Nonetheless, it is worth noting that as early as 1947 it was recognized \cite{segal1947} that the general state of a quantum system is a density matrix and that the ensemble interpretation is not necessary. 

It is interesting therefore to observe that stochastic reduction models can be readily formulated, as we have shown here, with the property that the status of the density matrix as representing the general state of a single isolated quantum system is sustained. In fact, the theory assumes a simpler form when it is pursued at the level of the density matrix, as we have seen, and thus forms a satisfactory basis upon which one can pursue more general classes of models such as the L\'evy models we consider later.  
%
%%%%%%%%%%%%%%%%%%%%%%%%
\section{Diagonalization of the density matrix}
\label{sec:3}
%%%%%%%%%%%%%%%%%%%%%%%%
From a purely probabilistic stance one can nevertheless give a completely consistent interpretation of the mean density matrix 
\begin{eqnarray}
{\hat\mu}_t = {\mathbb E}\left[ {\hat\rho}_t \right]
\end{eqnarray}
in terms of ensembles. One envisages a collection of $N$ independent identical copies of the given quantum system, each evolving from the same initial state $\hat \rho_0$. In each case the evolution is governed by an equation of the form \eqref{density matrix dynamics}, but with an independent $\mathbb P$-Brownian driver. It follows by Kolmogorov's strong law of large numbers that as the number of systems gets large, the ensemble average of the density matrices converges almost surely to the expectation of the random density matrix arising in the case of a single system. Thus, if 
$\{\hat \rho^{\,r}_t\}_{r = 1, \dots , N}$ are the density matrix processes of the $N$ independent quantum systems, then
\begin{eqnarray}
\mathbb P \left( \lim_{N \to \infty} \, \frac{1}{N} \sum_{r =1}^N \hat \rho^{\,r}_t = {\hat\mu}_t \right) = 1.
\end{eqnarray}

It should be emphasized that in a stochastic model of the type we are considering, there is no metaphysics involved in giving an ensemble interpretation to the mean density matrix. Rather, as we have seen, such an interpretation follows rigorously as a consequence of the law of large numbers. Indeed, all valid statements about ensembles in the present context can be formulated in such terms. The law of large numbers is a theorem, not an intuition. 

As an example, it will be useful to consider the phenomenon of decoherence from this point of view. We shall show that starting from an arbitrary initial density matrix,  the mean density matrix diagonalizes in the frame of the energy projectors as $t$ approaches infinity. First we observe, as we remarked earlier, that as a consequence of \eqref{density matrix dynamics} we obtain the usual deterministic master equation for the mean density matrix:
\begin{eqnarray}
\frac{\rd{\hat\mu}_t}{\rd t} = -{\rm i} \hbar^{-1}[{\hat H},{\hat\mu}_t] +
\tfrac{1}{4} \sigma^2 {\mathcal L}_{\hat H}\,{\hat\mu}_t .
\end{eqnarray}
This can be derived by integrating equation \eqref{density matrix dynamics} with the introduction of the initial condition, then taking the expectation of each side of the equation, then using the Fubini theorem to interchange the integrals and the expectations, then differentiating with respect to $t$. A calculation \cite{abbh2001} shows that the solution of the deterministic master equation is 
\begin{eqnarray}
\hat\mu_t = \sum_j \hat \Pi_j\, {\hat\rho_0} \,\hat \Pi_j + \sum_{j \neq k}
\re^ { -{\rm i} \hbar^{-1} (E_j - E_k)t - \frac{1}{8}\sigma^2 (E_j-E_k)^2 t}  \,\hat \Pi_j \, {\hat\rho_0} \,\hat \Pi_k .  
\end{eqnarray}
One sees that the second term is exponentially damped as time passes and that asymptotically one is left with the first term alone, in which the mean density matrix is diagonalized. 

Thus one can say that asymptotically the ensemble is equivalent to that of a mixture of energy eigenstates of the projector type, where the proportion belonging to eigenvalue $E_j$ is given by ${\rm tr} \, (\hat \rho_{0} \, \hat \Pi_j )$. This is indeed the result one would expect on intuitive grounds in the statistical interpretation of standard quantum mechanics when $\hat \mu_{\infty}$ represents the state of a system after an energy measurement has been made but before the outcome is known. We should stress, however, that in the present context we are deriving the result directly from the stochastic dynamics of the independent elements of the ensemble, without assuming the statistical interpretation of standard quantum mechanics. 

%%%%%%%%%%%%%%%%%%%%%%
\section{Signal detection and state reduction}
\label{sec:4}
%%%%%%

Before we develop our theory of state reduction models based on L\'evy noise, it will be useful first to outline how a solution to the dynamical equation
(\ref{stochastic Schrodinger equation}) can be obtained in the case of 
a pure state. A corresponding solution to the stochastic 
mixed-state evolution equation can then be formulated by analogy. 
A general closed-form solution to 
(\ref{stochastic Schrodinger equation}) was obtained in 
\cite{bh2002-2} and developed in greater detail in \cite{bh2006}. 
Our purpose 
in this section is to review the solution methodology, 
which then paves the way towards L\'evy generalizations. 

It turns out that a highly effective approach to solving (\ref{stochastic Schrodinger equation}) is 
by \textit{constructing} the solution explicitly using techniques of signal 
detection, rather than \textit{solving} the given differential equation. Once 
the solution is constructed, it is straightforward to show that it satisfies the 
differential equation that we intended to solve. 
Hence we begin with the consideration of a classical 
signal detection problem, leaving aside quantum theory for the moment. 

In signal detection, one is typically interested in inferring the true value of a signal, 
or message, given noisy observations. Let us assume that the 
unknown signal is represented by a fixed, time-independent random variable 
$H$ defined on a probability space $(\Omega,{\mathcal F},{\mathbb P})$ 
such that $H$ takes the value $E_j$ with the probability 
$p_j$, where $j=1,2,\ldots,n$. Now suppose that the signal is revealed continuously in time at a 
constant rate $\sigma$, but the signal is obscured by a Brownian noise 
$\{B_t\}_{t\geq0}$. Then the noisy observations of the signal 
can be modelled by an information-providing process $\{\xi_t\}_{t\geq0}$ that takes the form 
\begin{eqnarray}
\xi_t = \sigma H t + B_t . 
\label{eq:2} 
\end{eqnarray}
The task in signal detection at time $t$ is to determine the best estimate 
of the signal $H$ given the observed ``signal-plus-noise'' time series 
$\{\xi_s\}_{0 \leq s\leq t}$ up to that time. 
The notion of  ``best'' estimate evidently depends on the criterion used to judge the merits of the estimate, 
but for a wide range of reasonable criteria, such as minimization of the 
quadratic error, the best estimate of $H$ is  the 
conditional expectation 
\begin{eqnarray}
H_t = \sum_{j=1}^n E_j \, {\mathbb P}(H=E_j\,|\,\{\xi_s\}_{0\leq s\leq t}) .  
\end{eqnarray} 
Thus we need to work out the conditional probabilities 
\begin{eqnarray} \label{conditional probabilities}
\pi_{jt} = {\mathbb P}(H=E_j\,|\,\{\xi_s\}_{0\leq s\leq t}), \quad j = 1, 2, \dots, n.
\end{eqnarray} 
Since the information process has the
Markov property (one can prove this), and since 
\begin{eqnarray} 
\lim_{t\to\infty} \frac{1}{\sigma t} \, \xi_t = H, 
\end{eqnarray}
one finds that \eqref{conditional probabilities} reduces to a simpler expression, namely, 
\begin{eqnarray} 
\pi_{jt} = {\mathbb P}(H=E_j \,|\, \xi_t),
\end{eqnarray}
which can be worked out explicitly by use of the following form of the Bayes formula: 
\begin{eqnarray} 
{\mathbb P}(H=E_j \,|\,\xi_t) = \frac{{\mathbb P}(H=E_j)\,\rho(\xi_t\,|\,H=E_j)}
{\sum_{k=1}^n {\mathbb P}(H=E_k)\,\rho(\xi_t\,|\,H=E_k)}  .
\end{eqnarray}
Here $\rho(x \,| \,H=E_j)$ denotes the value at $x \in \mathbb R$ of the conditional density function of the random variable 
$\xi_t$. By use of ${\mathbb P}(H=E_j)=p_i$, and the fact that conditional on 
$H=E_j$ the random variable $\xi_t$ is normally distributed with mean 
$\sigma E_j t$ and variance $t$, we deduce that 
\begin{eqnarray} 
\pi_{jt} = \frac{p_j\,\exp\left(\sigma E_j\, \xi_t - \frac{1}{2} \sigma^2 E_j^2 t\right) }
{\sum_{k=1}^n p_k\,\exp\left(\sigma E_k\, \xi_t - \frac{1}{2} \sigma^2 E_k^2 t\right) } , 
\label{eq:6} 
\end{eqnarray}
from which the best estimate of the signal can be determined. 

Let us work out the stochastic differential of the process $\{\pi_{jt}\}_{t\geq 0}$. By use of Ito's formula one finds that 
\begin{eqnarray} 
\rd\pi_{it} = \sigma \big( E_i - H_t \big) \pi_{it} 
\left( \rd\xi_t - \sigma H_t \, \rd t \right).
\label{eq:7} 
\end{eqnarray}   
Then if we define a process $\{W_t\}_{t \geq 0}$ by setting
\begin{eqnarray}
W_t = \xi_t - \sigma \int_0^t H_s \, \rd s   , 
\label{eq:28} 
\end{eqnarray} 
it can be shown \cite{bh2002-2, bh2006} that $\{W_t\}$ is a standard 
Brownian motion under ${\mathbb P}$. That is to say,  $\{W_t\}$ turns out to be an $\{\mathcal F_t\}$-adapted 
Gaussian process with mean zero and autocovariance ${\rm Cov}(W_s, W_t) = s$ for $0 \leq s \leq t$, with stationary and independent increments. 
Then we have 
\begin{eqnarray}
\rd\pi_{it} = \sigma \big( E_i - H_t \big) \pi_{it} \,\rd W_t. 
\end{eqnarray}
Furthermore, if we consider 
the square-root probability processes $\sqrt{\pi_{jt}}$, for $j=1,2,\ldots,n$,  then by use of Ito's 
lemma and (\ref{eq:7}) we deduce that 
\begin{eqnarray}
\rd\sqrt{\pi_{jt}} = \textstyle{\frac{1}{2}} \sigma \big( E_j - H_t \big) 
\, \sqrt{\pi_{jt}} \, \rd W_t - \textstyle{\frac{1}{8}} \sigma^2 
\big( E_j - H_t \big)^2 \sqrt{\pi_{jt}} \, \rd t.
\label{eq:9} 
 \end{eqnarray} 

With these results at hand, let us consider a quantum system characterized by a Hamiltonian ${\hat H}$ that may or may not be degenerate. We assume, for the moment, that the initial state of the system  is pure, with state vector $|\psi_0\rangle$. 
As before, we let $\hat \Pi_j$ denote the projection operator onto the Hilbert subspace associated to
the energy eigenvalue $E_j$. Let us denote by $|E_j\rangle$ the normalized L\"uders state obtained by projecting 
the initial state $|\psi_0\rangle$
onto the Hilbert subspace with energy eigenvalue $E_j$. Thus,
\begin{eqnarray} \label{Luders eigenstates}
|E_j\rangle = \frac{1} {\sqrt{p_j} }\, \hat \Pi_j |\psi_0\rangle, \quad p_j = \frac { \langle \psi_0 | \hat \Pi_j |\psi_0\rangle} {\langle \psi_0 | \psi_0\rangle}.
\end{eqnarray} 
Next, we define a state vector process $\{|\psi_t\rangle\}_{t\geq 0}$ by setting
\begin{eqnarray} \label{eq:10}
|\psi_t\rangle = \sum_{j} \sqrt{\pi_{it}} \, \re^{-{{\rm i} \hbar^{-1}}E_jt} \, 
|E_j\rangle, 
\end{eqnarray} 
where the $\pi_{jt}$ are given by (\ref{eq:6}). 
Then a calculation making use of (\ref{eq:9}) shows that 
$|\psi_t\rangle$ is a solution to the stochastic 
Schr\"odinger equation (\ref{stochastic Schrodinger equation}) with the initial 
condition $|\psi_0\rangle$.
The advantage 
of the filtering method is that one can work directly with the 
solutions of the stochastic differential equation. The construction of  (\ref{eq:10}), along with (\ref{eq:6}), only 
requires the computation of the conditional probabilities. In particular, no stochastic
integration is required to arrive at the solution. This is the approach that we shall use shortly when we turn to look at collapse models based on L\'evy information.

%%%%%%%%%%%%%%%%%%%%%%%%
\section{Solution to the stochastic master equation}
\label{sec:5}
%%%%%%%%%%%%%%%%%%%%%%%%

The structure of the solution to the stochastic Schr\"odinger equation obtained in the previous section sheds light on foundational issues and at the same time suggests generalizations from 
a mathematical perspective. 
In particular, in the case of the stochastic master equation 
(\ref{density matrix dynamics}), a solution for the density matrix
can be constructed by the same line of argument. 

In the setting of a system based on finite-dimensional Hilbert space we regard the Hamiltonian $\hat H$ (possibly degenerate) and the initial state $\hat \rho_0$  (possibly of low rank) as being given. 
As before, let us write ${\hat \Pi}_j$ $(j = 1, \dots, n )$ for the projection operator onto the eigenspace of energy $E_j$. Then we fix a probability space $(\Omega, \mathcal F, \mathbb P)$ upon which we define a Brownian motion $\{B_t\}_{t\geq 0}$ along with an independent random variable $H$ taking values in the set $(E_j)_{j = 1, \dots, n} $ such that 
\begin{eqnarray} \label{probability assignment}
{\mathbb P}(H=E_j)={\rm
tr}({\hat\rho}_0 \, {\hat \Pi}_j ).
\end{eqnarray}
Next, we introduce a Brownian information process of the form (\ref{eq:2}). 
Finally, we set
\begin{eqnarray} \label{Krauss}
{\hat K}_t = \sum_{j=1}^n {\hat \Pi}_j  \, \re^{{\rm i}\hbar^{-1}E_j t + \frac{1}{2}
\sigma E_j \xi_t - \frac{1}{4} \sigma^2 E_j^2 t} .
\end{eqnarray}
Then one can show that the solution for the state process $ \{ {\hat\rho}_t\}_{t \geq0} $ is given by
\begin{eqnarray}\label{Brownian solution}
{\hat\rho}_t=\frac{{\hat K}_t^\dagger {\hat\rho}_0 {\hat K}_t}
{{\rm tr}({\hat K}_t^\dagger {\hat\rho}_0 {\hat K}_t)},
\end{eqnarray}
where the information process $\{\xi_t\}_{t \geq0} $ is related to the Brownian 
driver $\{W_t\}_{t \geq0} $ of (\ref{density matrix dynamics}) in accordance with  (\ref{eq:28}). 
In particular, one can prove that $ \{ {\hat\rho}_t\}_{t \geq0} $ satisfies the dynamical equation \eqref{density matrix dynamics} with  the Brownian driver 
$\{W_t\}_{t \geq0}$ and the prescribed initial condition and that $ \{ {\hat\rho}_t\}_{t \geq0} $ has the properties (a), (b), (c), (d), (e) stated in Section \ref{sec:2}. 

%%%%%%%%%%%%%%%%
\section{L\'evy Information}
\label{sec:6}
%%%%%%%%%%%%%%%%

We turn to a generalization of the foregoing considerations to a much wider class of processes. Let us fix a probability space $(\Omega, \mathcal F, \mathbb P)$ on which we define a L\'evy  process $\{\xi_t\}_{t\geq 0}$. By a L\'evy  process we mean a random process with stationary, independent increments. 
Brownian motion is an example of a L\'evy  process and indeed it is the only example of a continuous L\'evy  process, i.e.~a process with continuous 
sample paths. One can think of the different types of L\'evy  processes 
as representing different types of homogeneous noise. 

We shall assume in the following that $\{\xi_t\}_{t\geq 0}$ admits exponential moments. 
By this we mean that  there exists an open interval $S \subset \mathbb R$ containing the origin such that $S \subset C$ where
\begin{eqnarray}\label{exponential moments condition}
C = \left\{ c\in{\mathds R}:\,\, {\mathbb E}\left[\exp{c\,\xi_t}\right]  < \infty \right\} .
\end{eqnarray}
It can be shown that for any L\'evy  process admitting exponential moments there exists a strictly convex function $\psi: C \to \mathbb R$ such that
\begin{eqnarray}
\frac{1}{t} \log {\mathbb E}[\exp{\alpha \xi_t}] = \psi(\alpha)
\end{eqnarray}
for $\alpha \in C$. We refer to $\psi$ as the L\'evy exponent associated with the  L\'evy  process $\{\xi_t\}_{t\geq 0}$. 
By the L\'evy-Khintchine theorem \cite{applebaum, sato}, which is one of the foundational results of the theory,
there exists a constant $p$, a constant $q \geq 0$, and a 
L\'evy measure $\nu(\rd x)$ such that
\begin{eqnarray} \label{LK representation}
\psi(\alpha) = p\alpha + \half q \alpha^2 + \int_{{\mathbb R} }
\big(\re^{\alpha z} -1 -\alpha z {\mathds 1}\{|z|<1\}\big) \nu(\rd z) ,
\end{eqnarray}
and we refer to \eqref{LK representation} as the L\'evy-Khintchine representation. By a L\'evy measure $\nu$ on $\mathbb R$ we mean a $\sigma$-finite (but not necessarily finite)
measure  satisfying $\nu(\{0\}) = 0$ and 
\begin{eqnarray} \label{Levy measure}
 \int_{{\mathbb R} } 
\min(1, z^2) \, \nu(\rd z) < \infty .
\end{eqnarray}
We call
$\{p, q, \nu(\rd z)\}$ the characteristic triplet of the L\'evy process. Note that $\nu(A)$ is finite on any interval $A \in \mathbb R$ bounded away from the origin, but may be infinite on an interval that includes the origin.  
If a L\'evy process has L\'evy measure $\nu(\rd x)$, the rate at which jumps arrive for which the jump size is in the interval $[a, b]$ for $a < b$ with $\{ 0 \} \not \in [a, b]$ is given by 
\begin{eqnarray}
m[a,b] = \int_{[a, b]} \nu(\rd z),
\end{eqnarray}
which by \eqref{Levy measure} is evidently finite. If $a > 0$ and $\lim_{a \to 0} m[a, b] = \infty$ or if  
$b < 0$ and $\lim_{b \to 0} m[a, b] = \infty$ then we say that the L\'evy process has infinite activity. In this case, the process admits infinitely many very small jumps in any finite interval of time. 
Otherwise, the process has finite activity and can be represented by a compound Poisson process, in which case the normalized measure
\begin{eqnarray}\label{compound distribution}
p(\rd z) = \frac{\,\nu(\rd z)} { \int_{\mathbb R} \nu(\rd z) }
\end{eqnarray}
gives the probability distribution of the size of a typical jump, and jumps arrive at the rate
\begin{eqnarray} 
m_{\nu} =  \int_{\mathbb R} \nu(\rd z).
\end{eqnarray}

We are now in a position to define L\'evy information 
\cite{bhyang}. The idea of a L\'evy information process is that it generalizes 
the information process (\ref{eq:2}) for Brownian noise to the general class of
L\'evy processes introduced above. 
In the case of a signal obscured by Brownian noise, it 
is natural that the signal and noise should admit an additive decomposition. 
This is why we often hear the phrase ``signal plus noise'' in this context. The 
result is a Brownian motion $\{B_t\}$ with a linear drift, as we see in (\ref{eq:2}). One can 
then find a change of measure such that a drifted Brownian 
motion $\{\xi_t\}$ under the physical measure ${\mathbb P}$ becomes a pure 
Brownian motion under another probability measure, say, ${\mathbb P}^0$. 
This is the measure in which the observed information $\{\xi_t\}$ is content 
free -- that is to say, free of any signal $H$. The change of probability measure arising in this context, concerning which we shall have more to say in Section \ref{sec:7},  is called 
an Esscher transformation. 
Conversely, once the noise type (e.g., say, a Brownian noise) is specified, one can 
begin with the ``empty'' probability measure ${\mathbb P}^0$ in which the 
observation $\{\xi_t\}$ represents pure noise of the type selected, and then 
apply an Esscher transform to the physical measure ${\mathbb P}$ using the 
signal $H$. 

In this way, a type of information process can be created that can carry a 
much wider class of noise structures, not just Brownian noise. 
In particular, by 
letting $\{\xi_t\}$ be a ${\mathbb P}^0$-L\'evy noise, it is possible to naturally 
extend the theory of signal detection with Brownian noise into the general 
L\'evy setup introduced above. This procedure, in turn, allows one to formulate models of state reduction in 
quantum mechanics driven by a range of different L\'evy processes, each with its own special
characteristics. 

With these preliminaries at hand, let $\psi:C \to \mathbb R$ be a L\'evy 
exponent and let $X$ be a random variable taking values in $C$. 
Then by a L\'evy information process with information $X$ and L\'evy noise type $\psi$ we mean a process $\{\xi_t\}_{t\geq 0}$ that is conditionally L\'evy with a conditional exponent of the form
\begin{eqnarray} \label{conditional moment}
\frac{1}{t} \log {\mathbb E}[\exp{\alpha \xi_t} \, |\, \mathcal F_X] = \psi (\alpha + X)  -\psi(X),
\end{eqnarray}
where $\mathcal F_X$ is the $\sigma$-algebra generated by the random variable $X$. By conditionally 
L\'evy we mean that $\{\xi_t\}_{t\geq 0}$ has conditionally stationary and independent increments. 

One can show, for example, that the Brownian information process considered earlier satisfies these conditions. 
In the Brownian case we have a Gaussian exponent
\begin{equation}\label{Gaussian exponent}
\psi(\alpha) = p \alpha +  \half q \alpha^2
\end{equation}
 and $S = \mathbb R$, so we see that the conditional exponent  \eqref{conditional moment} in this case takes the form 
 \begin{eqnarray}
\psi (\alpha + X)  -\psi(X) = \alpha X + \half \alpha^2, 
\end{eqnarray}
with a random term  linear in $\alpha$. 
In fact, one can prove that a L\'evy information process can be constructed in association with any L\'evy process admitting exponential moments and any integrable random variable taking values in $C$. 
In the case of the Poisson process, for example, we have 
\begin{equation}
\psi(\alpha) = m (\re^\alpha - 1),
\end{equation}
where $m$ is the intensity, the rate at which the events being counted occur on average. 
For a Poisson information process the conditional moment takes the form 
\begin{eqnarray}
\psi (\alpha + X)  -\psi(X) = m\, \re^X (\re^\alpha - 1), 
\end{eqnarray}
showing that the intensity is randomized (or ``modulated") \cite 
{bhyang, sdk1975} and is given by 
$m\, \re^X $. 

Thus, in the case of Poisson noise, the information process 
$\{\xi_t\}$ is a Poisson process with intensity $m$ under the ``content 
free'' measure ${\mathbb P}^0$. 
But under the physical measure 
${\mathbb P}$ the process has a randomized intensity $m\,\re^{X}$. The 
observer therefore detects Poisson jumps, from which the task is to infer 
the jump intensity $m\,\re^{X}$, and hence the value of the 
signal $X$. The observer is already aware of the base rate $m$, and thus by counting the number of jumps over some interval of time they can estimate the value of $X$. 

The idea can be illustrated as follows. Imagine a situation where a laboratory may have been contaminated with a small amount of radioactive substance. A Geiger counter is used to measure the radiation level. If there is no contamination, the counter will click randomly at a low rate of activity, corresponding to the normal level of background radiation, but if the contaminant is present the counter will click at a higher rate. In this example, $X$ can take two possible values, with $X = 0$ corresponding to the normal rate of background activity and $X = \log (1 + \epsilon)$ for some $\epsilon > 0$ corresponding to the case where there is contamination. Analogously, for each type of L\'evy noise one can think of a class of signal detection problems, for which the available observations are represented by L\'evy information processes.

%%%%%%%%%%%%%%%%%%%%%%%%%%%%%%%
\section{Change of Measure}
\label{sec:7}
%%%%%%%%%%%%%%%%%%%%%%%%%%%%%%%
To establish the existence of L\'evy information processes we can
use a so-called change-of-measure technique.  Let us fix a 
probability space $(\Omega,{\mathcal F},{\mathbb P}^0)$ where 
${\mathbb P}^0$ will be called the base measure. We assume that 
$(\Omega,{\mathcal F},{\mathbb P}^0)$ supports a L\'evy process 
$\{\xi_t\}_{t\geq0}$ admitting exponential moments and we define the set $C$ as
in \eqref{exponential moments condition}.
Equivalently, 
\begin{eqnarray}
C = \left\{ c\in{\mathbb R}:\, 
\int_{\mathbb R} \re^{cz} \, \nu(\rd z) < \infty \right\}\!, 
\end{eqnarray}
where $\nu(\rd z)$ is the L\'evy measure associated with $\{\xi_t\}$. Now let $\{{\mathcal F}_t\}_{t\geq0}$ be the filtration 
generated by $\{\xi_t\}$. 
One 
can check that the process $\{\Lambda_t^\kappa\}_{t\geq 0}$ defined for 
$\kappa\in C$ by 
\begin{eqnarray} 
\Lambda_t^\kappa = \exp \left({\kappa \xi_t - \psi(\kappa)t} \right)
\end{eqnarray} 
is a martingale. This follows on account of the independent increments property of 
L\'evy processes. Then for each $t\in{\mathds R}^+$ we 
can define a measure ${\mathbb P}_t^\kappa$ on $(\Omega,{\mathcal F_t})$ 
by setting 
\begin{eqnarray}
{\mathbb P}_t^\kappa(A) = 
{\mathbb E}^{{\mathbb P}_t^{\kappa}} \!
\left[ {\mathds 1}(A) \right] 
={\mathbb E}^{{\mathbb P}^0} \!
\left[ \Lambda_t^\kappa \, {\mathds 1}(A) \right] 
\end{eqnarray}
for any $A\in{\mathcal F}_t$. The martingale property of $\{\Lambda_s^\kappa\}_{0 \leq s \leq t}$ ensures that 
${\mathbb P}_t^\kappa$ is a probability measure, since 
${\mathbb P}_t^\kappa(\Omega) = {\mathbb E}^{{\mathbb P}^0}\left[ \Lambda_t^\kappa \mathds 1 \{\Omega\}\right] = \Lambda_0^\kappa = 1$.
We observe that if $s\leq t$ and if $A$ is 
${\mathcal F}_s$-measurable, then 
\begin{eqnarray}
{\mathbb P}_t^\kappa(A) &=& {\mathbb E}^{{\mathbb P}^0} 
\left[ {\mathbb E}^{{\mathbb P}^0} 
\left[ \Lambda_t^\kappa \, {\mathds 1}(A) |{\mathcal F}_s \right] \right] 
\nonumber \\ &=& {\mathbb E}^{{\mathbb P}^0} 
\left[ {\mathbb E}^{{\mathbb P}^0} 
\left[ \Lambda_t^\kappa  |{\mathcal F}_s \right] \, {\mathds 1}(A) \right] 
\nonumber \\ &=&  {\mathbb E}^{{\mathbb P}^0} 
\left[ \Lambda_s^\kappa \, {\mathds 1}(A) \right]  \nonumber \\ &=&  
{\mathbb P}_s^\kappa(A), 
\end{eqnarray}
which shows that the measures defined on ${\mathcal F}_s$ and 
${\mathcal F}_t$ are compatible for $s\leq t$, in the sense that if 
$A\in{\mathcal F}_s$ then the measure of $A$ on $(\Omega,{\mathcal F}_s,
{\mathbb P}_s^\kappa)$ is the same as its measure on $(\Omega,
{\mathcal F}_t,{\mathbb P}_t^\kappa)$. With that in  mind,  we can ease the notation by dropping the 
subscript $t$ on ${\mathbb P}_t^\kappa$. 

When $\{\xi_t\}$, 
which is by assumption a L\'evy process on $(\Omega,{\mathcal F}, 
{\mathbb P}^0)$, is restricted to the time frame $\{\xi_s\}_{0\leq s\leq T}$, then it 
is also a L\'evy process on $(\Omega,{\mathcal F_T}, {\mathbb P}^{\kappa})$, for any 
choice of $T$. That is to say, it can be shown that for each $T\geq 0$ the process $\{\xi_s\}_{0\leq s\leq T}$ has stationary and independent increments under ${\mathbb P}^{\kappa}$.

But when $\{\xi_s\}_{0\leq s\leq T}$ is regarded as a process on $(\Omega,{\mathcal F_T}, 
{\mathbb P}^{\kappa})$, its properties shift:  if $\{p_0,q_0, 
\nu_0(\rd z)\}$ is the characteristic triplet of $\{\xi_t\}_{0\leq s\leq T}$ when it is regarded as 
a L\'evy process on $(\Omega,{\mathcal F}, {\mathbb P}^0)$, then on 
$(\Omega,{\mathcal F}, {\mathbb P}^\kappa)$ the process has a transformed characteristic 
triplet $\{p_{\kappa},q_\kappa, \nu_\kappa(\rd z)\}$ of the form
\begin{align}
p_\kappa = p_0 + \kappa q_0 + \int_{\mathbb R} (\re^{\kappa z}-1) 
{\mathds 1}(|z|<1)\, z \, \nu(\rd z), \quad q_\kappa = q_0 , \quad 
 \nu_\kappa(\rd z) = \re^{\kappa z}\, \nu(\rd z)
\end{align} 
Such a shift is called an Esscher transformation \cite{esscher1932}. Thus, a L\'evy 
process, when viewed from the untransformed probability space, has 
different characteristics from those of the same process when it is viewed 
from the Esscher-transformed probability space. 

The idea of a \textit{L\'evy information process} involves a similar construction, 
where we randomize the parameter of the Esscher transformation. 
Let $(\Omega,{\mathcal F}, {\mathbb P}^0)$, $\{\xi_t\}_{t\geq 0}$, and 
$\{{\mathcal F}_t\}_{t\geq 0}$ be as above, and let $X$ be an integrable 
random variable such that $\{\xi_t\}$ and $X$ are ${\mathbb P}^0$-independent. 
Let $\{{\mathcal G}_t\}_{t\geq 0}$ be the filtration generated jointly by $\{\xi_t\}$ 
and $X$. Then for each $t\geq 0$ we have 
\begin{align}
{\mathcal G}_t=\sigma\left[ \{\xi_s\}_{0
\leq s\leq t}, X\right]\!, 
\end{align} 
and clearly it holds that ${\mathcal F}_t 
\subset {\mathcal G}_t$. 
The next step is to define a new probability measure ${\mathbb P}_t^X$ on 
${\mathcal G}_t$ by setting 
\begin{eqnarray} 
{\mathbb P}_t^X(A) = {\mathbb E}^{{\mathbb P}^0} \! 
\left[ \Lambda_t^X \, {\mathds 1}(A) \right] ,  \quad \Lambda_t^X = \re^{X\xi_t - \psi(X) t} , 
\end{eqnarray}
for any $A \in {\mathcal G}_t$. It is straightforward to check that the process $\{\Lambda_s^X\}_{0\leq s 
\leq t}$ is a martingale on $(\Omega,{\mathcal G}_t, {\mathbb P}^0)$, 
which ensures that the measures $\{{\mathbb P}_t^X(A)\}_{t\in{\mathds R^+}}$ 
are compatible for various values of $t$, so we can drop the $t$  
and write ${\mathbb P}^X(A)$ for the transformed measure. 
To proceed further we need the following formula
for the conditional expectation with respect to ${\mathcal F}_t$. 

%==========
%Proposition 1
%==========
\begin{Proposition}\label{prop:1}
For any integrable random variable $Z$ on 
$(\Omega,{\mathcal G}_t, {\mathbb P}^X)$ it holds that 
\begin{eqnarray}
{\mathbb E}^{{\mathbb P}^X}\! \left[ Z\,|\,{\mathcal F}_t\right] = 
\frac{{\mathbb E}^{{\mathbb P}^0}\! \left[ \re^{X\xi_t-\psi(X)t} \, 
Z\,|\,{\mathcal F}_t\right] }{{\mathbb E}^{{\mathbb P}^0} \! \left[ 
\re^{X\xi_t-\psi(X)t} \,|\,{\mathcal F}_t\right] } . 
\label{eq:KS} 
\end{eqnarray} 
\end{Proposition}

\noindent \textit{Proof}. We recall Kolmogorov's definition of conditional expectation, given in Section \ref{sec:1}. 
In the 
present situation the role of $Y$ is played by the right side of (\ref{eq:KS}), 
so we must show that 
\begin{eqnarray}
{\mathbb E}^{{\mathbb P}^X}\! \left[ Y\,{\mathds 1}(A)\right] = 
{\mathbb E}^{{\mathbb P}^X}\! \left[ Z\,{\mathds 1}(A)\right] 
\label{Kolmogorov condition} 
\end{eqnarray} 
for any ${\cal F}_t$-measurable set $A$, where 
\begin{eqnarray}
Y = 
\frac{{\mathbb E}^{{\mathbb P}^0}\! \left[ \re^{X\xi_t-\psi(X)t} \, 
Z\,|\,{\mathcal F}_t\right] }{{\mathbb E}^{{\mathbb P}^0} \! \left[ 
\re^{X\xi_t-\psi(X)t} \,|\,{\mathcal F}_t\right] } . 
\end{eqnarray} 
But if $A$ is ${\cal F}_t$-measurable we 
obtain 
\begin{eqnarray}
{\mathbb E}^{{\mathbb P}^X}\! \left[ Y\,{\mathds 1}(A)\right]   &=&
 {\mathbb E}^{{\mathbb P}^0}\! \left[ \re^{X\xi_t-\psi(X)t} \, 
 \frac{ {\mathbb E}^{{\mathbb P}^0}\! \left[ \re^{X\xi_t-\psi(X)t} \, 
Z \,|\,{\mathcal F}_t\right] }
   {{\mathbb E}^{{\mathbb P}^0} \! \left[ 
\re^{X\xi_t-\psi(X)t} \,|\,{\mathcal F}_t\right] }  {\mathds 1}(A) \right], 
\nonumber \\ &=& 
{\mathbb E}^{{\mathbb P}^0}\! \left[ \re^{X\xi_t-\psi(X)t} \, 
 \frac{ {\mathbb E}^{{\mathbb P}^0}\! \left[ \re^{X\xi_t-\psi(X)t} \, 
Z {\mathds 1}(A)\,|\,{\mathcal F}_t\right] }
   {{\mathbb E}^{{\mathbb P}^0} \! \left[ 
\re^{X\xi_t-\psi(X)t}\,|\,{\mathcal F}_t\right] } \right], 
\nonumber \\ &=& 
{\mathbb E}^{{\mathbb P}^0}\! \left[ {\mathbb E}^{{\mathbb P}^0}\! \left[ \re^{X\xi_t-\psi(X)t} \, 
 \frac{   {\mathbb E}^{{\mathbb P}^0}\! \left[ \re^{X\xi_t-\psi(X)t} \, 
Z {\mathds 1}(A)\, |\,{\mathcal F}_t\right] }
   {  {\mathbb E}^{{\mathbb P}^0} \! \left[ 
\re^{X\xi_t-\psi(X)t}\, |\,{\mathcal F}_t\right] }\, |\,{\mathcal F}_t  \right] \right]
\nonumber \\ &=& 
 {\mathbb E}^{{\mathbb P}^0}\! \left[  {\mathbb E}^{{\mathbb P}^0}\! \left[ \re^{X\xi_t-\psi(X)t} \, 
Z {\mathds 1}(A)\,|\,{\mathcal F}_t\right]  \right]
\nonumber \\ &=& 
{\mathbb E}^{{\mathbb P}^X}\! \left[ Z\,{\mathds 1}(A)\right]\!,  
\end{eqnarray} 
and thus we deduce (\ref{Kolmogorov condition}) by repeated use of the tower property. 
\hfill $\Box$
\vspace{0.5cm}

\noindent In fact, \eqref{eq:KS} arises naturally as a form of the so-called Kallianpur-Striebel formula \cite{KS}.  A special case of \eqref{eq:KS}  is particularly useful. 
%==========
%Proposition 2
%==========
\begin{Proposition} \label{prop:2}
Let the function $f:\, {\mathds R} \to {\mathds R}$ be such that $f(X)$ is integrable. Then we have  
\begin{eqnarray}
{\mathbb E}^{{\mathbb P}^X}\! \left[ f(X)|{\mathcal F}_t\right] = 
\frac{ \int f(x)\, \re^{x\xi_t-\psi(x)t} \mu(\rd x) }
{\int \re^{x\xi_t-\psi(x)t} \mu(\rd x) } ,
\label{eq:KS2} 
\end{eqnarray} 
where the distribution of $X$ is given by
\begin{eqnarray}
{\mathbb P}(X\leq a) = \int_{-\infty}^a \mu(\rd x) . 
\end{eqnarray} 
\end{Proposition}

Now suppose that $\{\xi_t\}_{t\geq0}$ is a L\'evy process on 
$(\Omega,{\mathcal F}, {\mathbb P}^0)$ with L\'evy exponent $\psi(\alpha)$ 
for $\alpha\in{C}$. Then, for any times $t$ and $T$ such that $0 \leq t \leq  T$ we have
\begin{eqnarray}
{\mathbb E}^{{\mathbb P}^X}\!\left[ \re^{\alpha\xi_t}\, |\, {\cal F}_X\right] 
&=& {\mathbb E}^{{\mathbb P}^0}\!\left[ \re^{X\xi_t-\psi(X)t} 
\re^{\alpha\xi_t} \,|\, {\cal F}_X\right] \nonumber \\ &=& 
{\mathbb E}^{{\mathbb P}^0}\!\left[ \re^{(X+\alpha)\xi_t-\psi(X)t}\,|\,{\cal F}_X\right] 
\nonumber \\ &=& \re^{(\psi(X+\alpha)-\psi(X))t} ,
\end{eqnarray} 
since $X$ and $\{\xi_t\}_{0 \leq t\leq T}$ are independent on  $(\Omega, 
{\mathcal F}_T, {\mathbb P}^0)$, and this gives \eqref{conditional moment}. A similar calculation shows that $\{\xi_t\}_{t\geq0}$ has ${\cal F}_X$-conditionally stationary and independent increments under $\mathbb P^X$.  We thus deduce that 
$\{\xi_t\}_{t\geq0}$ is a L\'evy information process on $(\Omega,{\mathcal F}, {\mathbb P}^X)$ and hence that L\'evy information processes can be constructed for any noise type.
%
%%%%%%%%%%%%%%%%%%%%%%%%%%%%%%%
\section{Quantum State Reduction with L\'evy Jumps}
\label{sec:8}
%%%%%%%%%%%%%%%%%%%%%%%%%%%%%%%
%
Suppose that we are given a quantum system defined on a finite dimensional Hilbert space with Hamiltonian $\hat H$ and initial state $\hat \rho_0$. 
We fix a probability space $(\Omega, \mathcal F, \mathbb P)$ on which we define a random variable $H$ taking values in the set 
$\{E_j\}_{j = 1, \dots, n}$  such that \eqref{probability assignment} holds. 
We define a L\'evy information process $\{\xi_t\}_{t\geq 0}$ carrying the information of $H$ such that
\begin{eqnarray} \label{conditional exponent}
\frac{1}{t} \log {\mathbb E}[\exp{\alpha \xi_t} \, |\, \mathcal F_H] = \psi (\alpha + \lambda H)  -\psi(\lambda H),
\end{eqnarray}
where $\lambda$ is a model parameter with the units 
\begin{eqnarray} 
\lambda \sim[{\rm
energy}]^{-1}.
\end{eqnarray}
Our conventions going forward are such that the L\'evy information process is dimensionless, as is the argument of the L\'evy exponent. Thus $H$ has units of energy, $t$ has units of time, and the L\'evy exponent has units of inverse time.   

Now consider the situation where the initial state is pure, so  $\hat \rho_0 = |\psi_0\rangle \langle \psi_0 |$ for some given initial state vector $|\psi_0\rangle$. Then for the dynamics of a L\'evy-driven 
state vector  leading to a collapse of the wave function to an energy 
eigenstates we generalize the approach laid out in Section \ref{sec:4} and look at a model of the form 
\begin{eqnarray} 
|\psi_t\rangle = \sum_{j=1}^n \sqrt{\pi_{jt}} \, \re^{-{\rm i \hbar^{-1}}E_jt} \, 
|E_j\rangle, \quad  \pi_{jt} = {\mathbb E}[\mathds 1 \{H=E_j\} \,|\,\mathcal F_t], 
\end{eqnarray}
where the $|E_j\rangle$ are the normalized L\"uders eigenstates given by \eqref{Luders eigenstates} and $\pi_{jt}$ is the conditional probability  for reduction to a state with energy $E_j$, given L\'evy information. Then we apply Proposition \ref{prop:2} in the case for which
the distribution of the random variable $X = \lambda H$ is 
\begin{eqnarray} \label{formula for measure}
\mu(\rd x) = \sum_{j=1}^n p_j \, \delta_{\lambda E_j} (\rd x),
\end{eqnarray}
where $\delta_{y} (\rd x)$ denotes the Dirac measure concentrated at $y$, and we obtain
\begin{eqnarray} 
\pi_{jt} = \frac{p_j\,\exp\left(\lambda E_j\, \xi_t - \psi(\lambda E_j) t\right) }
{\sum_{k=1}^n p_k\,\exp\left(\lambda E_k\, \xi_t - \psi(\lambda E_k) t\right) }  
\end{eqnarray} 
for the conditional probabilities. That gives us our model for the collapse dynamics of a state vector driven by L\'evy information. 

More generally, for the dynamics of a L\'evy-driven 
density matrix leading to a reduction of the initial state to an energy 
eigenstate, we propose a model of the form 
\begin{eqnarray} \label{Levy solution}
{\hat\rho}_t = \frac{\re^{-{\rm i} \hbar^{-1} \hat H t+ \frac{1}{2}\lambda{\hat H} \xi_t  
- \frac{1}{2} \psi(\lambda{\hat H})t}\,{\hat\rho}_0
\,\re^{{\rm i}\hbar^{-1} \hat H t+ \frac{1}{2}\lambda{\hat H} \xi_t  - \frac{1}{2} \psi(\lambda{\hat H})t}}
{{\rm tr}\left({\hat\rho}_0\, \re^{\lambda{\hat H} \xi_t  -\psi(\lambda{\hat H})t} \right)} . 
\end{eqnarray}
It is straightforward to check that when ${\hat\rho}_0$ is pure,  our model for the dynamics of the density matrix is consistent with the state vector dynamics considered above. One can also check that the associated mean density matrix satisfies a deterministic dynamical equation of the Lindblad type. This is far from obvious but the proof will be given in Section \ref{sec:9}. Then in Section \ref{sec:10} we show that the model leads to reduction of the state to an energy eigenstate of the L\"uders type satisfying the Born rule.
 
One can see at a glance that \eqref{Levy solution} reduces to an expression of the form \eqref{Brownian solution}  in the Brownian case, but the precise relation may not be immediately obvious, since the units of the parameter $\sigma$ are not the same as those of $\lambda$. In the Brownian situation it is convenient to choose units such that $B_t$ has dimensions of square-root time, in which case  the information process defined by $\sigma H t + B_t$ likewise has units of square-root time, which implies that the units of the parameter $\sigma$ are given by \eqref{units}.  In the Brownian case such choices are convenient and widely used. 

If, however, one wishes to treat Brownian motion as a species of L\'evy process alongside and in additive combination with other L\'evy processes,  one needs a single convention that traverses the L\'evy category. This means that L\'evy processes should be made dimensionless. 

After all, there are many examples of 
L\'evy processes that take the form of counting processes, which are dimensionless in their natural setting. 
The dimensionless Brownian motion with drift associated with the Gaussian exponent \eqref{Gaussian exponent}
takes the form $p t + q^{1/2} B_t $, where $\{B_t \}_{t\geq 0}$ is a standard Brownian motion (with units of square-root time), and the corresponding dimensionless L\'evy information process $\{\xi_t \}_{t\geq 0}$ with conditional exponent \eqref{conditional exponent} is given by
\begin{eqnarray} \label{Brownian dimensionless}
\xi_t =  q \lambda H t + q^{1/2} B_t .
\end{eqnarray} 
The Brownian information process introduce in Section \ref{sec:4} is obtained by dividing \eqref{Brownian dimensionless} by $q^{1/2}$ and setting 
$\sigma = q^{1/2} \lambda$, which gives  \eqref{units}. This may seem little complicated, but there is a clash between the conventions of physicists, who treat Brownian motion as if it has the physical units of square-root time, and probability theorists, who regard Brownian motion (and time itself) as dimensionless. As long as we work with Brownian motion alone, either convention will do, but once general L\'evy processes are brought into play, we find that the probabilistic conventions work better.

A compromise can be reached by choosing the second as the unit of time and treating $q$ as a model parameter. Then if we choose $q = 1 \,{\rm Hz}$ we recover the formulae that we used earlier for a standard Brownian information process with parameter $\sigma$. But in general we need a flexible value for $q$, since it characterizes the weighting of the Gaussian component of a L\'evy process relative to its other components. 

In fact, once adjustments are made to take into account the units, one can show that the main results mentioned earlier in connection with the Brownian model go through for a L\'evy model.
Hence from a physical point of view, we reach the important conclusion that there is no obvious reason {\em a priori} to prefer 
a Brownian model over any other L\'evy model. 

%%%%%%%%%%%%%%%%%%%%%%%%%%%%%%%
\section{Calculation of the decoherence rate}
\label{sec:9}
%%%%%%%%%%%%%%%%%%%%%%%%%%%%%%%

Now consider  the mean density matrix 
${\hat\mu}_t = {\mathbb E}\left[ {\hat\rho}_t \right]$. Here the expectation is calculated  under the physical measure and it can be worked out explicitly using Proposition 1 and a change-of-measure trick. In particular, writing $\mathbb P^0$ for the base measure, we  use the tower property and the fact that $\xi_t$ is $F_t$-measurable to show that
\begin{eqnarray}
{\mathbb E}^{\mathbb P}\!\left[ \hat \rho_t \right] 
&=& 
{\mathbb E}^{{\mathbb P}^0}\!\left[ \re^{\lambda H\xi_t-\psi(\lambda H)t} \,
\hat \rho_t \right] \nonumber \\ 
&=& 
 {\mathbb E}^{{\mathbb P}^0}\!\left[{\mathbb E}^{{\mathbb P}^0}\!\left[ \re^{\lambda H \xi_t-\psi(\lambda H)t} \hat \rho_t \,|\,{\cal F}_t\right] \right]
\nonumber \\ 
&=& 
{\mathbb E}^{{\mathbb P}^0}\!\left[ \int_{\mathbb R} \re^{\lambda x\xi_t-\psi(\lambda x)t} \mu(\rd x) \,
\hat \rho_t \right] \!,
\end{eqnarray} 
where $\mu(\rd x)$ is the probability measure of $H$. Now, by \eqref{probability assignment} and \eqref{formula for measure} we have
\begin{eqnarray}
\int_{\mathbb R} \re^{\lambda x\xi_t-\psi(\lambda x)t} \mu(\rd x) = \sum_{k=1}^n p_k\,\exp\left(\lambda E_k\, \xi_t - \psi(\lambda E_k) t\right)
= {\rm tr}\left({\hat\rho}_0\, \re^{\lambda{\hat H} \xi_t  -\psi(\lambda{\hat H})t} \right)\!, 
\end{eqnarray} 
since $\hat H = \sum_{j=1}^n E_j \hat \Pi_j$. The term in the denominator of  \eqref{Levy solution} gets cancelled and we obtain
\begin{eqnarray}
{\mathbb E}^{\mathbb P}\!\left[ \hat \rho_t \right] = 
{\mathbb E}^{\mathbb P^0}\!\left[ \re^{-{\rm i} \hbar^{-1} \hat H t+ \frac{1}{2}\lambda{\hat H} \xi_t  - \frac{1}{2} \psi(\lambda{\hat H})t}\,{\hat\rho}_0
\,\re^{{\rm i}\hbar^{-1} \hat H t+ \frac{1}{2}\lambda{\hat H} \xi_t  - \frac{1}{2} \psi(\lambda{\hat H})t} \right]\!. 
\end{eqnarray} 
Because
the exponential moments of $\xi_t$ under $\mathbb P^0$ can be expressed in terms of the L\'evy exponent, we are able to work out the following exact expression for the 
matrix elements of $\hat \rho_t$ with respect to the energy eigenspace projectors:
\begin{eqnarray}
\hat \Pi_m \, {\hat\mu_t} \,\hat \Pi_n = 
\re^ { -{\rm i} \hbar^{-1} (E_m - E_n)t + \psi\left(\frac{1}{2}\lambda (E_m+E_n)\right)t-\frac{1}{2} 
\psi(\lambda E_m)t-\frac{1}{2} \psi(\lambda E_n)t}  \,\hat \Pi_m \, {\hat\mu_0} \,\hat \Pi_n .  
\end{eqnarray}
Thus, more succinctly, we can write 
\begin{eqnarray}\label{matrix elements}
\hat \Pi_m \, {\hat\mu_t} \,\hat \Pi_n = 
\exp \left( -{\rm i} \hbar^{-1} (E_m - E_n)\,t - \Gamma_{mn} \,t \right)  \,\hat \Pi_m \, {\hat\mu_0} \,\hat \Pi_n ,  
\end{eqnarray}
where
\begin{eqnarray} 
 \Gamma_{mn}  = \half \psi(\lambda E_m)+ \half \psi(\lambda E_n) -\psi\left(\half \lambda (E_m+E_n)\right)\!.
\end{eqnarray}
Then as a consequence of the L\'evy-Khintchine representation \eqref{LK representation}
we deduce that 
\begin{eqnarray} \label{decoherence rate}
\Gamma_{mn} = \octa q \lambda^2(E_m-E_n)^2 + \half 
\int_{{\mathds R} } \left( \re^{\frac{1}{2}\lambda E_m z} - 
 \re^{\frac{1}{2}\lambda E_n z} \right)^2 \nu(\rd z) .
\end{eqnarray}
The key point is that $\Gamma_{mn}$ vanishes along the diagonal and is strictly positive for $m \not = n$. This positivity can also be seen to follow from the fact that the L\'evy exponent is strictly convex. We conclude that {\em the mean density matrix diagonalizes as $t$ gets large}.  This is the decoherence effect induced by the reduction process.

The dynamical equation satisfied by the mean density matrix can be worked out by differentiating \eqref{matrix elements} and takes the form
\begin{eqnarray}
\frac{\rd{\hat\mu}_t}{\rd t} &=& \ri [{\hat H},{\hat\mu}_t] +
\quat q \lambda^2 \left( {\hat H} {\hat\mu}_t {\hat H} -\half {\hat\mu}_t 
{\hat H}^2 - \half {\hat H}^2 {\hat\mu}_t \right) \nonumber \\ && 
+  \int_{-\infty}^\infty \left( {\hat L}(z) {\hat\mu}_t {\hat L}(z) - \half
{\hat L}^2(z) {\hat\mu}_{t} - \half {\hat\mu}_{t} {\hat L}^2(z) \right) \nu(\rd z) , 
\end{eqnarray}
which is evidently of the Lindblad type, where
\begin{eqnarray}
{\hat L}(z) = \re^{\frac{1}{2}{\lambda \hat H} z} . 
\end{eqnarray}
This is consistent with the idea that if a stochastic modification of standard quantum mechanics is to avoid causality violation it must lead to Lindblad-type dynamics for the mean density matrix 
\cite{gisin1989, gisin1990, polchinski1990, weinberg2012, adler2016}, and we see that this condition is satisfied by the class of models presently under consideration.  

Indeed, as \cite{adler2016} puts the matter, the overall structure of an objective reduction model is fixed by two natural physical requirements: ``The first is the requirement of state vector normalization -- the unit norm of the state vector should be maintained in time. The second is the requirement that there should be no faster than light signaling -- the density matrix averaged over the noise should satisfy a linear evolution equation of Lindblad form."

Some insight into the nature of the decoherence process can be gained if we analyze the terms of \eqref{decoherence rate}. 
Let us consider first the Brownian case, which just involves the first term of \eqref{decoherence rate}. In this case we can set
$\sigma^2 = q \lambda^2$ and we find that the decoherence rate is given by 
\begin{eqnarray} 
\Gamma_{mn} = \octa \sigma^2(E_m-E_n)^2 
\end{eqnarray}
for the matrix element corresponding to a typical pair of energy levels, where $\sigma$ has units of the form \eqref{units}. This result for the decoherence rate is consistent with rule of thumb that the reduction time scale in the Brownian case goes like 
$
\tau_R \sim 1 / \sigma^2 \, V_0 ,
$
as we mentioned earlier in Section \ref{sec:1}, where $V_0$ denotes the initial value of the squared uncertainty in the energy. For clearly, $V_0$ can be expressed as a weighted sum of squares of differences of energy eigenvalues. For example, for a two-level system one has $V_0 = p_1 p_2 (E_1 - E_2)^2$,  in the case of a three-level systems one has 
$V_0 = p_1 p_2 (E_1 - E_2)^2 +  p_1 p_3 (E_1 - E_3)^2 + p_2 p_3 (E_2 - E_3)^2$, and so on. 

The phenomenological case for energy-based reduction with a Brownian noise has been investigated extensively \cite{hughston1996, ah2000, adler2002, adler2003, pearle2004, adler2016}. If one takes the view that state reduction is linked in some way to gravitation, as many have, then a reasonable guess for $\sigma$ based on dimensional analysis is that it is given by a relation of the form 
\begin{eqnarray} 
\sigma^2 \sim  M_{p}^{-2} \, T_p^{-1}  = {2.8} \,{ \rm MeV}^{-2} s^{-1}. 
\label{Planck} 
\end{eqnarray}

An interesting feature of this conjecture is that the large numbers of the Planck mass $M_p$ and the Planck time $T_p$ cancel, and one is left with a laboratory-scale value for the reduction parameter \cite{hughston1996}. The data show that such a value for $\sigma$ is not unreasonable, in the sense that none of the many situations that have been analyzed in detail rule it out decidedly, although there is little by way of direct evidence in favour of it, at least as matters stand. 

As an example we can look at the framework proposed by Weinberg \cite{weinberg2016} in his analysis of the decoherence timescales associated with experiments involving atomic clocks. He derives a Lindblad equation based on his approach and argues that the resulting decoherence rate $\Gamma$ coming from objective reduction must satisfy a bound of the form $\Gamma T < 1$ where $T$ is the Ramsey time associated with the clock. He concludes that, ``Unfortunately we have no idea of what target of $\Gamma$ we should aim at, or even how $\Gamma$ might vary from one transition to another." Weinberg then proceeds to examine two extreme cases, namely where $\Gamma$ is constant and where $\Gamma$ scales like $E_e - E_g$ where $E_e$ and $E_g$ denote the excited and ground energy levels of the stable states of the clock. In the case of a ${^{133} \rm Cs^+}$ ion, for example, the hyperfine transition frequency is known to great accuracy and is given by $\Delta \nu_{\rm Cs} = 9.192\,631\,77$ GHz. This is of course the definition of the Hertz. The corresponding energy difference is then $\Delta E_{\rm Cs}  = h \Delta \nu_{\rm Cs} \approx 6.091 \times 10^{-17} \,{\rm erg}$ or equivalently 
$3.801 \times 10^{-5} \,{\rm eV}$. The Ramsey time varies according to the type of clock but is typically of the order of a  few seconds although Weinberg points to a case involving ${^{171} \rm Yb^+}$ for which 
$T > 600 \, {\rm s}$. Arguing on this basis he concludes that $\Gamma < 10^{-18} {\rm eV}$ if $\Gamma$ does not depend on the transition frequency.  

In the case of an energy-based reduction model with Brownian noise one can take matters a step further, because the precise dependence of $\Gamma$ on the transition frequency is given by equation  \eqref{decoherence rate} in that model. This allows us to work out a bound on $\sigma$. To get a feeling for the numbers involved, let us consider the ${^{133} \rm Cs}^+$ hyperfine frequency and use a Ramsey time of $1\, {\rm s}$. Then by  \eqref{decoherence rate} we have 
\begin{eqnarray}
\Gamma_{eg} = \octa \sigma^2 (\Delta E_{\rm Cs})^2 < 1 \, {\rm s}^{-1} 
\end{eqnarray}
and hence
$
 \sigma^2  <   {8}  \,(\Delta E_{\rm Cs})^{-2}   \, {\rm s}^{-1} 
$
which gives
$
 \sigma^2  <   0.5537 \times 10^{22}   \, {\rm MeV}^{-2}   \, {\rm s^{-1} }.
$
Thus we obtain an upper bound on  $\sigma^2$ from the atomic clock data, though not a particularly stringent one, and certainly the Planckian value for $\sigma^2$ that we considered in \eqref{Planck} is well within it. 

This example illustrates the fact that at the current level of technology it is not easy to identify decisive tests that would rule out energy-based reduction models based on Brownian noise, even though such tests are clearly possible in principle. 
Let us now turn to the L\'evy case. Here there are some surprises. Suppose for simplicity we consider pure jump models for which $|\lambda E_j z| \ll 1$ for all jump sizes in the support of the L\'evy measure. In that case we can make a Taylor expansion of the exponential terms inside the integral with respect to the L\'evy measure in our formula \eqref{decoherence rate} for the decoherence rate, neglecting the Brownian terms, and we obtain
\begin{eqnarray} 
\Gamma_{mn} = \octa  \lambda^2(E_m-E_n)^2 
\int_{{\mathds R} } z^2 \, \nu(\rd z) .
\end{eqnarray}
The surprising thing here is that when this approximation is valid (i.e.~for small jumps, small energies, small $\lambda$) the
expression for the decoherence rate is of the same form as that of the Brownian model, with the second moment of the L\'evy measure playing the role of the $q$ parameter. Thus, as long as the effects of the L\'evy model are in some sense perturbative, they do not qualitatively change the conclusions of the Brownian reduction models. 

We proceed to another surprise. Weinberg \cite{weinberg2016} appears in his analysis of atomic clocks simply to have assumed that decoherence rate depends on the transition frequencies, but not the overall levels of the energy eigenvalues. This is in certain respects a plausible assumption, and indeed it holds in Brownian reduction models. But in the L\'evy reduction models the situation is different. In general, the decoherence rate depends on the  overall levels of the energy eigenvalues as well as on their differences. This can be seen if we write 
\eqref{decoherence rate} as
\begin{eqnarray} \label{sinh formula}
\Gamma_{mn} = 2\!
\int_{{\mathds R} }  \re^{\frac{1}{2} \lambda (E_m + E_n) z} 
 \sinh^2 \left( \quat \lambda (E_m - E_n) z  \right) \nu(\rd z) ,
\end{eqnarray}
again neglecting the Brownian terms. For simplicity, suppose we consider the case of a spectrally positive L\'evy process (positive jumps) in the situation where the energy levels are non-negative. In that case, the decoherence rate is clearly an increasing function of $E_m + E_n$. Then even in the situation where the energy gaps are small, i.e.~such that  $\lambda | E_m - E_n\,| z  \ll 1$, one will get a high rate of decoherence if the overall energy levels are high, satisfying $\lambda ( E_m + E_n) z  \gg 1$. Thus, large systems will decohere quickly, even if the transition frequencies are small, whereas small systems will not decohere. 

In this respect the L\'evy models differ fundamentally from their Brownian counterparts. 
In particular, noting that $\sinh x \sim x$ for $x\ll 1$ we see that \eqref {sinh formula} reduces to
\begin{eqnarray} \label{integral for rate} 
\Gamma_{mn} \approx \octa  \lambda^2(E_m-E_n)^2 \!
\int_{{\mathds R} } \re^{\frac{1}{2} \lambda (E_m + E_n) z}  z^2 \, \nu(\rd z) ,
\end{eqnarray}
similar to the Gaussian case in its dependence on the transition frequency, but the effect of the large energies is to enhance the effective value of the $q$ parameter. For example, in the case of a Poisson process of intensity $m_{\nu}$, for which the jumps are of size unity, the L\'evy measure is given by $\nu(\rd z) = m_{\nu} \delta_1(\rd z)$ and we obtain
\begin{eqnarray} 
\Gamma_{mn} \approx \octa m_{\nu} \re^{\frac{1}{2} \lambda(E_m + E_n)} 
\lambda^2(E_m-E_n)^2 ,   
\end{eqnarray}
 for which the effective $q$ factor takes the form
 \begin{eqnarray} 
\tilde q_{mn} =  m_{\nu} \,
 \re^{\frac{1}{2} \lambda(E_m + E_n)}.
  \end{eqnarray}
 This example shows how the decoherence rate
 increases exponentially in the Poisson model when the overall energy scale is increased. 
 
 In fact, the integral on the right hand side of equation \eqref{integral for rate} can be computed explicitly in terms of the L\'evy exponent and we get
\begin{eqnarray} 
\Gamma_{mn} \approx \octa  \lambda^2(E_m-E_n)^2 \,
\psi'' (\half \lambda (E_m + E_n) ) ,
\end{eqnarray}
 which is positive on account of the convexity of the L\'evy exponent. Thus, we have
 \begin{eqnarray} 
 \tilde q_{mn} = 
 \psi'' (\half \lambda (E_m + E_n) ),
 \end{eqnarray}
 which can be worked out explicitly in the case of various examples. Note that $\psi''(\alpha)$ is an increasing function of its argument if the L\'evy process is spectrally positive. 
 
In the case of a compound Poisson process, for which $\psi(\alpha) = m_{\nu} (\phi(\alpha) - 1)$, where $ \phi(\alpha)$ is the moment generating function of the random jump size, we obtain  $\tilde q = m_{\nu} \,\phi''(\half \lambda (E_m + E_n))$. 
 As a specific example of such a process, with spectral positivity, we look at the situation for which the jumps are exponentially distributed, with probability density
 $\mathbb P [\, Y \in \rd y\, ] = \beta \,\re^{-\beta y} \,\rd y $ for $y > 0$, with $\beta > 0$. The moment generating function is 
\begin{eqnarray}
 \mathbb E [\re^ {\alpha Y} ] = \frac {\beta}  {\beta - \alpha }, 
 \end{eqnarray}
with $\alpha < \beta$. Thus,  for $\half \lambda (E_m + E_n)  < \beta$ we deduce that the effective $q$ factor takes the form
 \begin{eqnarray} 
 \tilde q_{mn} = m_{\nu} \frac {2\beta}{\left(  \beta - \half \lambda (E_m + E_n) \right)^3} 
 \end{eqnarray}
in the case of a compound Poisson process with exponentially 
distributed jumps.  
As another example, we can consider the case of a gamma information process, which is applicable in the situation in which the energy eigenvalues satisfy $0 < \lambda E_m < 1$, the L\'evy exponent is given by the expression $\psi(\alpha) = - m_{\nu} \log(1-\alpha)$, for $0 < \alpha < 1$, and therefore
 \begin{eqnarray} 
 \tilde q_{mn} = m_{\nu} \frac {1}{\left( 1 - \half \lambda (E_m + E_n) \right)^2}.
 \end{eqnarray}

Finally, we remark on the implications of this analysis for the measurement problem. If measurement takes the form of making an entanglement of the measuring apparatus with the system being measured, then in the case of a
spectrally positive L\'evy model 
there is no need to invoke the idea that large-scale macroscopic superpositions are required for the outcome of the measuring apparatus. The coupling between the system and the apparatus can be such that the different possible outcomes for the system are linked to rather small differences in the overall energy of the apparatus. The state of the apparatus will collapse nonetheless, despite the energy differences being small, on account of the amplification effect we have just discussed, bringing with it a collapse of the state of the small system to which it is coupled. This may explain why, within the Copenhagen interpretation of standard quantum mechanics, the mere act of measuring the energy of a small system forces it into an eigenstate. Likewise our approach may explain why the formation of a latent image in a photographic emulsion may be sufficient to collapse the state of the system being measured even if the image is developed at a later time \cite{gisinpercival1993, adler2018}.

%%%%%%%%%%%%%%%%%%%%%%%%%%%%%%%
\section{Proof of Reduction}
\label{sec:10}
%%%%%%%%%%%%%%%%%%%%%%%%%%%%%%%
To show that the L\'evy-information based models we have introduced in the previous sections have the reduction property, we establish a result that characterizes the asymptotic properties of the exponential martingale associated with a L\'evy process. 
%=============
%Proposition 3
%=============
\begin{Proposition} \label{prop:3}
%%%%%%%%%
Let $\{\xi_t\}_{t \geq 0}$ be a L\'evy process that admits exponential moments. 
Let $S$ denote the largest open set in ${\mathds R}$ such that 
\begin{eqnarray}
{\mathbb E}\left[ \re^{\kappa\xi_t}\right] < \infty 
\end{eqnarray}
for $\kappa\in S$. Then for any $\epsilon>0$ and any $\kappa\in S$ such that $\kappa \neq 0$ it holds 
that 
\begin{eqnarray}
\lim_{t\to\infty} {\mathbb P}\left( \re^{\kappa\xi_t-\psi(\kappa)t}
>\epsilon\right) =0 . 
\end{eqnarray}
\end{Proposition} 

\noindent \textit{Proof}. We require Cantelli's inequality \cite{Cantelli, Ghosh}, which is a strengthened version of Chebyshev's inequality holding in the case of a one-sided probability distribution that says that  for any 
square-integrable random variable $Z:~\Omega\to{\mathds R}$ and any $b\geq0$ it holds that 
\begin{eqnarray}
{\mathbb P}\left( Z-{\mathbb E}[Z]\geq b \right) \leq \frac{{\rm Var}[Z]}
{{\rm Var}[Z]+b^2}  \, .
\end{eqnarray}
Now, for any L\'evy process admitting exponential moments we have ${\mathbb E}[\xi_t]=
\psi'(0)t$ and ${\rm Var}[\xi_t]=\psi''(0)t$, from which by use of Cantelli's inequality we obtain
\begin{eqnarray}
{\mathbb P}\left( \re^{\kappa\xi_t-\psi(\kappa)t}>\epsilon\right) &=& 
{\mathbb P}\left( \xi_t> \kappa^{-1} [ \, \log\epsilon+\psi(\kappa) \, t \, ]\, \right) 
\nonumber \\ &=& 
{\mathbb P}\left( \xi_t-{\mathbb E}[\xi_t]> \kappa^{-1} [ \, \log\epsilon+(\psi(\kappa) - \kappa \psi'(0) )\, t \, ]\,  \right) \nonumber \\ &\leq& 
\frac{\psi''(0)t} {\psi''(0)t+  \kappa^{-2}\,[\,\log\epsilon+ ( \psi(\kappa)- \kappa \psi'(0))\,t \, ]^2} . 
\end{eqnarray}

\noindent But the convexity of the L\'evy exponent implies that $\psi(\kappa) > \kappa \psi'(0)$ for all $\kappa \in S - \{0\}$, and the claimed result follows immediately.  \qed 
\vspace{0.3cm}

Then to show that the  dynamical state process \eqref{Levy solution} reduces to an energy eigenstate it suffices to establish the following. 

%==========
%Proposition 4
\begin{Proposition} 
%==========
If the outcome of chance $\omega \in \Omega$ is such that $H(\omega) = E_j$ for some particular value of $j$, then for any $\epsilon > 0$ it holds that
\begin{eqnarray}
\lim_{t\to\infty} \mathbb P \big(1 - {\rm tr} \, ( \hat \Pi_{j} \hat \rho_t ) > \epsilon \big) = 0 . 
\end{eqnarray}
\end{Proposition} 

\noindent \textit{Proof}. It follows as a consequence of \eqref{Levy solution} and the cyclic property of the trace that
\begin{eqnarray}\label{conditional prob calculation}
{\rm tr}( \hat \Pi_{j} \hat \rho_t ) =  \frac { \re^{\lambda E_j \xi_t-\psi(\lambda E_j)t}  \, {\rm tr} \, ( \hat \Pi_{j} \hat \rho_0 ) }  
 { \sum _i \re^{\lambda E_i \xi_t-\psi(\lambda E_i)t}  \, {\rm tr} \, ( \hat \Pi_{i} \hat \rho_0 )  } ,  
\end{eqnarray}
and hence
\begin{eqnarray}
{\rm tr}( \hat \Pi_{j} \hat \rho_t ) =  \frac { {\rm tr} \, ( \hat \Pi_{j} \hat \rho_0 ) }  
 { {\rm tr} \, ( \hat \Pi_{j} \hat \rho_0 ) + \sum _{i\neq j} \re^{\lambda (E_i - E_j) \xi_t-\psi(\lambda E_i)t +\psi(\lambda E_j)t}  \, {\rm tr} \, ( \hat \Pi_{i} \hat \rho_0 )  } .  
\end{eqnarray}
Now, conditional on information $H = E_j$,  the process $\{\xi_t\}_{t \geq 0}$ is L\'evy, with L\'evy exponent 
$\psi (\alpha + \lambda E_j) - \psi ( \lambda E_j)$. It follows that the process $\{M_{ijt}\}_{t \geq 0}$ defined for $i \neq j$ by
\begin{eqnarray}
M_{ijt} =  \re^{\lambda (E_i - E_j) \xi_t-\psi(\lambda E_i)t +\psi(\lambda E_j)t} 
\end{eqnarray}
is an exponential martingale. We know therefore that $\{M_{ijt}\}$ converges to zero by Proposition \ref{prop:3}, from which we deduce that  
${\rm tr}\,\hat \Pi_{j} \hat \rho_t$ converges to unity.  \qed
\vspace{0.5cm}

Thus we have shown that the dynamical model defined by \eqref{Levy solution} is a state reduction process that carries the initial state to an energy eigenstate in such a way that the associated energy expectation process is a martingale. Indeed, the processes obtained by transvecting the state with any of the energy projection operators and taking the trace are likewise martingales. 
It follows that the actual probability of collapse to a L\"uders eigenstate of energy $E_j$  agrees with the probability calculated via the Born rule. Here we refer to the extended form of the Born rule fully applicable to density matrices and a possibly degenerate spectrum for the Hamiltonian. Results of this type have been known for some time in the case of Brownian noise 
\cite{gisin1989, gpr1990, hughston1996, ah2000, abbh2001, bh2002-2, bh2006, bh2018}, both for dynamics of state vectors and the dynamics of density matrices, but the extension to the L\'evy class is new. 

The new degrees of freedom that can be expressed in a L\'evy model are embodied in the structure of the L\'evy exponent for the underlying noise, or equivalently the parameter $q$ together with the L\'evy measure. In a number of situations one can construct explicit models of L\'evy information processes 
\cite{bhyang}. In such cases we can go further and use the model as a basis for simulation studies. In particular, in addition to those based on Brownian noise and Poisson noise, explicit models can be constructed for information processes based on L\'evy processes with infinite activity, including various examples that are well known in the theory of finance and insurance, such as (a) the gamma process \cite{dickson waters, BHM2008dam, yor} and (b) the variance gamma process \cite{Madan Milne 1991, BHMackie2012}. Such infinite activity information processes are different in character from their Brownian and Poisson counterparts.

In closing, we comment that our explicit formula for the decoherence rate \eqref{decoherence rate} marks a clear distinction between quantum state reduction models based on Brownian noise and the more general category of reduction models based on L\'evy noise, and paves the way towards possible applications of such models, some of which we have touched on in the present paper. The idea that changing the nature of the underlying noise might give some new insights into the measurement problem comes as a surprise and we hope to pursue the topic further. 

\begin{acknowledgments}
\noindent
We are grateful for comments from participants at the conference General Relativity, Quantum Mechanics and Everything in Between, Celebrating 92 Springs of Professor Lawrence Paul Horwitz, April 2022, Ariel University, Israel, where this work was presented. We thank Christopher Fuchs, Blake Stacey, and other seminar participants at the Department of Physics, University of Massachusetts Boston, for stimulating discussions in May 2022. We also thank Lajos Di\'osi, Philip Pearle and the anonymous referees for their comments. DCB acknowledges support from the Russian Science Foundation (grant 20-11-20226) and the Templeton Foundation (grant 62210). The opinions expressed in this publication are those of the authors and do not necessarily reflect the views of the Templeton Foundation.
\end{acknowledgments}

\vspace{0.5cm}
\newpage

\noindent {\bf APPENDIX: Examples of L\'evy Reduction Models} 

\vspace{0.5cm} 
It may be worthwhile if we explain things in more detail with a few concrete examples of L\'evy reduction models to show the occurrence of wave function collapse systematically. We look at reduction models for three canonical examples of L\'evy processes: (a) Brownian motion, (b) the Poisson process, and (c) the gamma process, these being representative of the continuous case, the finite activity case, and the infinite activity case. The Brownian model is of course well studied in the literature, but by presenting this example in the same notation and in parallel with the two other cases, we hope that the models based on pure jump L\'evy processes will be clearer. For simplicity we consider a quantum system with only two energy levels $E_1$ and $E_2$. The generalization to situations with more than two energy levels is straightforward and can be left as an exercise. 

Throughout the discussion that follows we are given a quantum system based on a finite dimensional Hilbert space with initial state $\hat \rho_0$ and a Hamiltonian operator with two energy levels. For the projectors on to the Hilbert subspaces with energies 
$E_1$ and $E_2$ respectively we write $\hat \Pi_{1}$ and $ \hat\Pi_{2}$. We consider a system for which  reduction is driven by a L\'evy information process $\{\xi_t\}_{t\geq 0}$ based on an underlying L\'evy noise with L\'evy exponent $\psi(\alpha)$, satisfying 
\begin{eqnarray} \label{conditional exponent2}
\frac{1}{t} \log {\mathbb E}[\exp{\alpha \xi_t} \, |\, \mathcal F_H] = \psi (\alpha + \lambda H)  -\psi(\lambda H),
\end{eqnarray}
where $\lambda$ is a model parameter and $H$ is a random variable taking the value $E_1$ with probability $ p_1 = {\rm tr} ( \hat \Pi_{1} \hat \rho_0)$ and the value $E_2$ with probability $ p_2 = {\rm tr} (\hat \Pi_{2} \hat \rho_0)$.

\vspace{0.5cm}

\noindent {\bf (a) Brownian noise.} 
We consider a dimensionless information process of the form $\xi_t =  q \lambda H t + q^{1/2} B_t $, where the parameter $q$ has dimensions of inverse time and  $\{B_t\}_{t\geq0}$ is a standard Brownian motion, with vanishing drift and variance $t$. For simplicity we set $q = 1 {\rm Hz}$ and then the information process takes the form 
\begin{eqnarray} \label{Brownian information process}
\xi_t =   \lambda H t +  B_t .
\end{eqnarray} 
The L\'evy exponent is given in the Brownian case by 
\begin{eqnarray}
\psi(\alpha) = \half \alpha^2 
\end{eqnarray} 
for $\alpha \in \mathbb R$. A straightforward calculation shows that for the conditional L\'evy exponent in this case we have 
\begin{eqnarray} 
\frac{1}{t} \log {\mathbb E}\left[\exp{\alpha  (\lambda H t +  B_t )} \, |\, \mathcal F_H\right] =
 \alpha  \lambda H  + \half \alpha^2 = \half (\alpha + \lambda H)^2  - \half(\lambda H)^2,
\end{eqnarray}
confirming that \eqref{Brownian information process} is an information process, satisfying \eqref{conditional exponent2}. 
The exponential martingale $\{M^\kappa_t\}_{t\geq0}$ with parameter $\kappa$ associated with a standard Brownian motion takes the form
\begin{eqnarray}
M^\kappa_t =  \re^{\kappa B_t- \frac{1}{2} \kappa^2 t}.
\end{eqnarray}
Such processes are widely used in the theory of finance to model the random fluctuations of share prices.  
Since $B_t$ is normally distributed with mean $0$ and variance $t$, then  for any value of $t > 0$, no matter how large, clearly we have 
\begin{eqnarray}
\mathbb E \left[\, \re^{\kappa B_t- \frac{1}{2} \kappa^2 t} \,\right] = 1. 
\end{eqnarray}
It may then come as a surprise that for any $\epsilon > 0$, no matter how small, we have
\begin{eqnarray}
\lim_{t \to \infty} \mathbb P  \left [ \re^{\kappa B_t- \frac{1}{2} \kappa^2 t}  > \epsilon \right ] = 0. 
\end{eqnarray}
That is to say, the exponential Brownian motion process converges to zero in probability. This can be shown with an application of Cantelli's inequality. 
For those who have any doubts, we can check the result directly by use of an old-school probability calculation. Writing $N(x)$, $x \in \mathbb R$, for the standard normal distribution function, we have 
\begin{eqnarray}
\mathbb P  \left [ \re^{\kappa B_t- \frac{1}{2}\kappa^2 t}  > \epsilon \right ] &=& 
\mathbb P  \left [ \frac {1} {t^{1/2}} B_t >  \frac{1}{ \kappa \,t^{1/2} }  \log \epsilon + \half \kappa t^{1/2} \right ] \nonumber \\ 
&=& 1 - N\left( \frac{1}{ \kappa \,t^{1/2} }  \log \epsilon + \half \kappa t^{1/2} \right)\!, 
\end{eqnarray}
since the random variable $t^{-1/2} B_t $ is normally distributed with mean zero and variance unity. But $\lim_{x \to \infty } N(x) = 1$ and the claimed result follows immediately.  Similar results hold for the exponential martingales associated with other L\'evy processes. 

Now we are in a position to look at the conditional probability for the outcome of the reduction process to be an eigenstate with
energy $E_1$ (the corresponding calculations for $E_2$ follow by symmetry). According to the general theory outlined in 
Section \ref{sec:8} we have
\begin{eqnarray}
{\rm tr}( \hat \Pi_{1} \hat \rho_t ) =  \frac {  p_1\re^{\lambda E_1 \xi_t-\frac{1}{2}\lambda^2  E_1^{\, 2}t} }  
 { p_1\re^{\lambda E_1 \xi_t-\frac{1}{2}\lambda^2  E_1^{\, 2}t}  + p_2 \re^{\lambda E_2 \xi_t-\frac{1}{2} \lambda^2  E_2^{\, 2}t}   } ,  
\end{eqnarray}
by virtue of equation \eqref{conditional prob calculation}, and hence
\begin{eqnarray}\label{probability of outcome one}
{\rm tr}( \hat \Pi_{1} \hat \rho_t ) =  \frac { p_1 }   
 {   p_1 + p_2  \re^{ \lambda (E_2 - E_1) \xi_t- \frac{1}{2}\lambda^2 E_2^{\, 2}t  + \frac{1}{2} \lambda^2  E_1^{\, 2}t}   } .  
\end{eqnarray}
To show that reduction occurs, we need to prove that if the outcome of chance is that $H = E_1$, then 
$\lim_{t \to \infty} {\rm tr}( \hat \Pi_{1} \hat \rho_t)  = 1$, whereas if the outcome of chance is $H = E_2$ then 
$\lim_{t \to \infty} {\rm tr}( \hat \Pi_{1} \hat \rho_t)  = 0$. This may not be obvious on a casual glance at 
\eqref{probability of outcome one}. But the point is that if $H = E_1$ then $\xi_t =   \lambda E_1 t +  B_t $. 
Substituting this into \eqref{probability of outcome one} we get
\begin{eqnarray}
\left. {\rm tr}( \hat \Pi_{1} \hat \rho_t )\right|_{H = E_1} =  \frac { p_1 }   
 {   p_1 + p_2  \re^{ \lambda (E_2 - E_1) (\lambda E_1 t +  B_t) - \frac{1}{2} \lambda^2 E_2^{\, 2}t  + \frac{1}{2} \lambda^2  E_1^{\, 2}t}   } ,  
\end{eqnarray}
and hence after some rearrangement, we have
\begin{eqnarray}
 {\rm tr}( \hat \Pi_{1} \hat \rho_t )\big |_{H = E_1} =  \frac { p_1 }   
 {   p_1 + p_2  \re^{ \lambda (E_2 - E_1) B_t - \frac{1}{2} \lambda^2 (E_2 - E_1)^{\, 2}t  }  } .  
\end{eqnarray}
We observe that an exponential martingale of the type we were considering earlier appears in the denominator. This converges to zero in probability, and it follows that for any $\epsilon > 0$ it holds that
\begin{eqnarray}\label{convergence property}
 \lim_{t \to \infty} \mathbb P  \left [ \left. {\rm tr}( \hat \Pi_{1} \hat \rho_t )\right|_{H = E_1} \!< 1-\epsilon\right]=  0 ,  
\end{eqnarray}
which signifies that reduction to a state of energy $E_1$ has taken place. A similar calculation then shows that
\begin{eqnarray}
\left. {\rm tr}( \hat \Pi_{1} \hat \rho_t )\right|_{H = E_2} =  \frac { p_1 }   
 {   p_1 + p_2  \re^{ \lambda (E_2 - E_1) B_t + \frac{1}{2}\lambda^2 (E_2 - E_1)^{\, 2}t  }  } ,  
\end{eqnarray}
and hence that
\begin{eqnarray}
 \lim_{t \to \infty} \mathbb P  \left [ \left. {\rm tr}( \hat \Pi_{1} \hat \rho_t )\right|_{H = E_2} > \epsilon \right] =  0 .  
\end{eqnarray}

\vspace{0.5cm}

\noindent {\bf (b) Poisson noise.} 
The Poisson process $\{ N_t\}_{t\geq 0}$ with intensity $m > 0$ is a nondecreasing jump process with unit jumps at the rate $m$.  The L\'evy exponent is given by
\begin{eqnarray}
\psi(\alpha) = m\, (\re^\alpha -1) 
\end{eqnarray} 
and the corresponding L\'evy measure takes the form 
\begin{eqnarray}
\nu(\rd z) = m\, \delta_1(\rd z),
\end{eqnarray} 
where $\delta_1(\rd z)$ denotes the Dirac measure concentrated at jump size unity. The Poisson process takes values in the nonnegative integers, whose distribution at time $t$ is 
\begin{eqnarray}
\mathbb P \left[ N_t = n \right] = \re^{-mt} \frac{\,\,(mt)^n} { n !}.
\end{eqnarray} 
For the corresponding exponential martingale with parameter $\kappa$ we have
\begin{eqnarray}\label{Poisson exponential martingale}
M^\kappa_t =  \re^{\kappa N_t- m ({\rm e}^\kappa -1)  t}.
\end{eqnarray}
A calculation gives  $\mathbb E [N_t ] = mt$ and $\Var[N_t ] = mt$. Cantelli's inequality then tells us that
\begin{eqnarray}
{\mathbb P}\left( \re^{\kappa\xi_t-\psi(\kappa)t}>\epsilon\right) \leq
\frac{mt} {mt+  \kappa^{-2}\,[\,\log\epsilon+ m( {\rm e}^\kappa -1 - \kappa)\,t \, ]^2} . 
\end{eqnarray}
One can check that $\inf_{\kappa \in \mathbb R} ({\rm e}^\kappa -1 - \kappa) = 0$ and hence that for any $\kappa > 0$ the exponential martingale  \eqref{Poisson exponential martingale} converges to zero in probability as $t$ grows large. 

A Poisson information process with parameter $m$ can be modelled by letting $\{N(t)\}_{t \geq 0}$ be a standard Poisson process with parameter $m$ as described above and setting
\begin{eqnarray}\label{Poisson information process}
\xi_t = N(\re^{\lambda H} t) .
\end{eqnarray}
For the conditional L\'evy exponent in this case we have 
\begin{eqnarray} 
\frac{1}{t} \log {\mathbb E}[\exp{\alpha  N(\re^{\lambda H} t)} \, |\, \mathcal F_H] =
m\,\re^{\lambda H} (\re^\alpha -1)    =  m\,(\re^{\alpha + \lambda H} -1)  - m\,(\re^{\lambda H} -1),
\end{eqnarray}
which shows that \eqref{Poisson information process} satisfies \eqref{conditional exponent2} and thus is indeed an information process. The conditional probability for reduction to an eigenstate with eigenvalue $E_1$ is 
\begin{eqnarray}
{\rm tr}( \hat \Pi_{1} \hat \rho_t )& = & \frac {  p_1\,\re^{\lambda E_1 \xi_t-m ({\rm e}^{\lambda E_1} -1)t} }  
 { p_1\,\re^{\lambda E_1 \xi_t-m ({\rm e}^{\lambda E_1} -1)t}  + p_2\,\re^{\lambda E_2 \xi_t-m ({\rm e}^{\lambda E_2} -1)t}   } \nonumber \\
 & = &  \frac {  p_1}  
 { p_1 + p_2\,\re^{\lambda (E_2-E_1) \xi_t-m {\rm e}^{\lambda E_1}({\rm e}^{\lambda (E_2- E_1)} -1)t}   } .
\end{eqnarray}
Therefore,
\begin{eqnarray}
 \left. {\rm tr}( \hat \Pi_{1} \hat \rho_t )\right|_{H = E_1}  =   \frac {  p_1}  
 { p_1 + p_2\,\re^{\lambda (E_2-E_1) \,\xi_t |_{H = E_1} -m {\rm e}^{\lambda E_1}({\rm e}^{\lambda (E_2- E_1)} -1)t}   } .  
\end{eqnarray}
However, $\xi_t |_{H = E_1}$ is a standard Poisson process with rate $m\,\re^{\lambda E_1}$, so an exponential martingale appears in the expression above, from which we get \eqref{convergence property} and hence reduction. 

\vspace{0.5cm}

\noindent {\bf (c) Gamma noise.} 
Physicists are familiar with Brownian motion and the Poisson process, both of which appear frequently in the literature. The gamma process is a newer idea, dating from the 1950s, exhibiting interesting features with which physicists are 
perhaps less familiar.  Most important among these is the phenomenon of {\em infinite activity}. 
By a  gamma process with rate $m$  and scale $\varphi$ we mean a L\'evy process 
$\{\gamma_t\}_{t \geq 0}$ with L\'evy exponent 
\begin{eqnarray}\label{gamma exponent}
\psi(\alpha) = 
-m \log (1 - \varphi \alpha) ,
\end{eqnarray}
where $\alpha  < \varphi^{-1}$. 
The probability density for $\gamma_t$ is that of the gamma distribution, given by 
\begin{eqnarray}
{\mathbb P}({\mathit\gamma}_t \in \rd x) = 
\frac{ \varphi ^{-mt} x^{mt-1} \re^{-x/\varphi } } { \Gamma[mt]} \, \rd x 
\end{eqnarray}
for $x > 0$, and zero otherwise, where $\Gamma[a]$ is the gamma function. A calculation using the identity $\Gamma[a + 1] = a \Gamma[a ]$ then shows that 
${\mathbb E}\,[\gamma_t] = m\varphi t $ and 
${\rm Var}\,[\gamma_t] = m\varphi^2 t. $ Note that the mean and variance 
determine the rate and scale. If $\varphi = 1$ then $\{\gamma_t\}$ is called a 
\textit{standard} gamma process with rate $m$. 
If $\varphi\neq1$ we say that $\{ \gamma_t\}$ is a \textit{scaled} gamma process.
The L\'evy measure associated with $\{\gamma_t\}_{t \geq 0}$ is given by
\begin{eqnarray}
\nu( {\rd} z) =  m\, \frac{1}{z}  \exp ({-\varphi z})\, {\rd} z
\end{eqnarray}
for $z > 0$, and zero otherwise. It follows that $\nu ({\mathds R} ) = \infty$ and hence that the gamma 
process has infinite activity.  Thus, the jumps are all positive and the number of jumps in any finite interval of time is infinite. 

Let $\{\gamma_t\}$ be a standard gamma process with rate  $m$ and let the parameter $\kappa \in \mathbb R$ be such that
$\kappa<1$. Then the process $\{M_t^\kappa\}$ defined by 
\begin{eqnarray} \label{gamma exponential martingale}
M_t^\kappa = (1-\kappa )^{mt} \, \re^{\kappa\gamma_t}
\end{eqnarray}
is a martingale and by Cantelli's inequality we have
\begin{eqnarray}
{\mathbb P}\left( (1-\kappa )^{mt} \, \re^{\kappa\gamma_t} >\epsilon \right) \leq
\frac{mt} {mt+  \kappa^{-2}\,[\,\log\epsilon+ m( - \log(1-\kappa) - \kappa)\,t \, ]^2} . 
\end{eqnarray}
One can check that $- \log(1-\kappa) - \kappa > 0$ for all $\kappa < 1$. This follows from the basic logarithmic inequality
$\log x \leq x - 1$  for all $x \geq 0$ with equality at $x = 1$. Hence for any $\kappa < 1$ the exponential gamma martingale  \eqref{gamma exponential martingale} converges to zero in probability as $t$ grows large. 

If we let $\{M_t^\kappa\}$ act as 
a change of measure density for the transformation ${\mathbb P}^0\to{\mathbb P}^\kappa$, 
then $\{\gamma_t\}$ is a \textit{scaled} gamma process under 
${\mathbb P}^\kappa$, with rate $m$ and scale $1/(1-\kappa)$. Thus, the effect 
of an Esscher transformation on a gamma process is to alter its scale. 

Now let $\{\gamma_t\}$ be a standard gamma process with rate $m$ and let the independent random 
variable $X$ satisfy $X<1$ almost surely. Then the process $\{\xi_t\}$ defined by
\begin{eqnarray}
\xi_t = \frac{1}{1-X}\, \gamma_t 
\label{gamma information process}
\end{eqnarray}
is a L\'evy information process with signal $X$ and gamma noise. 
Thus $\{\xi_t\}$ is conditionally a scaled gamma process. Then as a 
consequence of \eqref{gamma exponent} and \eqref{gamma information process} we have 
\begin{eqnarray}
\frac{1}{t} \ln {\mathbb E}^{\mathbb P}\left[\exp(\alpha \xi_t)|\mathcal F^X \right]  = 
\frac{1}{t} \ln {\mathbb E}^{\mathbb P}\left[\left.\exp\left(\frac{\alpha\gamma_t}{1-X}
\right) \right| \mathcal F^X\right]  = \log \left(1-\frac{\alpha}{1-X}\right)^{-m}\!. 
\end{eqnarray}
Next, we observe that
\begin{eqnarray}
-m \ln \left(1-\frac{\alpha}{1-X}\right) =
-m \ln \left(1-(X + \alpha) \right)  \,+\, m \ln \left(1-X \right)\!. 
\end{eqnarray}
It follows that the conditional exponent of $\{\xi_t\}$ takes the form 
$\psi(\alpha + X)-\psi(X)$, which shows that $\{\xi_t\}$ is an information process.

We proceed to look at the conditional probability for reduction to an energy eigenstate with energy $E_1$ in the context of a gamma information model. In this case the parameter $\lambda$ is chosen to satisfy $\max (\lambda E_1, \lambda E_2) < 1$
and we have
\begin{eqnarray}
{\rm tr}( \hat \Pi_{1} \hat \rho_t )& = & \frac {  p_1\,\re^{\lambda E_1 \xi_t } (1-\lambda E_1 )^{mt} }  
 { p_1\,\re^{\lambda E_1 \xi_t } (1-\lambda E_1 )^{mt}  + p_2\,\re^{\lambda E_2 \xi_t } (1-\lambda E_2 )^{mt}   } \nonumber \\
 & = &  \frac {  p_1}  
 { p_1 + p_2\,\re^{\lambda (E_2-E_1) \xi_t}  (1-\lambda E_2 )^{mt}  (1-\lambda E_1 )^{-mt}  } ,
\end{eqnarray}
and therefore
\begin{eqnarray}
\left. {\rm tr}( \hat \Pi_{1} \hat \rho_t )\right|_{H = E_1}  & = &   \frac {  p_1}  
 { p_1 + p_2\,\re^{\lambda (E_2-E_1) \xi_t}|_{H = E_1}  (1-\lambda E_2 )^{mt}  (1-\lambda E_1 )^{-mt}  }  \nonumber \\
 & = & \frac {  p_1}  
 { p_1 + p_2\,\re^{\lambda (E_2-E_1) (1-\lambda E_1 )^{-1}\gamma_t}  (1-\lambda E_2 )^{mt}  (1-\lambda E_1 )^{-mt}  }. \label{coefficient in gamma probability} 
\end{eqnarray}
However, 
\begin{eqnarray}
\left (1 - \frac{\lambda (E_2-E_1) } {1-\lambda E_1}  \right )^{mt} = (1-\lambda E_2 )^{mt}  (1-\lambda E_1 )^{-mt},
\end{eqnarray}
and therefore the second term in the denominator on the right side of \eqref{coefficient in gamma probability}
is an exponential gamma martingale, which converges to zero in probability for large $t$. Thus we have  \eqref{convergence property}.

\vspace{1cm}

\newpage
\noindent {\bf References}
%+Bibliography

%\section*{References}
%\bibliographystyle{apsrmp}
%\bibliographystyle{apalike}

%\bibliography{References}

\end{document}